# Superconductivity in pressurized trilayer La$_4$Ni$_3$O$_{10-\delta}$ single crystals

Yinghao Zhu[1,2†], Di Peng[3†], Enkang Zhang[1†], Bingying Pan[4†], Xu Chen[5†], Lixing Chen[1], Huifen Ren[5], Feiyang Liu[1], Yiqing Hao[6], Nana Li[7], Zhenfang Xing[7], Fujun Lan[7], Jiyuan Han[1], Junjie Wang[5,8], Donghan Jia[9], Hongliang Wo[1], Yiqing Gu[1], Yimeng Gu[1], Li Ji[10], Wenbin Wang[11], Huiyang Gou[9], Yao Shen[5], Tianping Ying[5], Xiaolong Chen[5], Wenge Yang[7], Huibo Cao[6], Changlin Zheng[1], Qiaoshi Zeng[3,7*], Jian-gang Guo[5*] & Jun Zhao[1,2,11,12*]

[1]*State Key Laboratory of Surface Physics and Department of Physics, Fudan University, Shanghai 200433, China*
[2]*Shanghai Research Center for Quantum Sciences, Shanghai 201315, China*
[3]*Shanghai Key Laboratory of Material Frontiers Research in Extreme Environments (MFree), Institute for Shanghai Advanced Research in Physical Sciences (SHARPS), Shanghai 201203, China*
[4]*College of Physics and Optoelectronic Engineering, Ocean University of China, Qingdao, Shandong 266100, China*
[5]*Beijing National Laboratory for Condensed Matter Physics, Institute of Physics, Chinese Academy of Sciences, Beijing 100190, China*
[6]*Neutron Scattering Division, Oak Ridge National Laboratory, Oak Ridge, Tennessee 37831, USA*
[7]*Center for High Pressure Science and Technology Advanced Research, Shanghai 201203, China*
[8]*School of Physical Sciences, University of Chinese Academy of Sciences, Beijing 100049, China*
[9]*Center for High Pressure Science and Technology Advanced Research, Beijing 100094, China*
[10]*State Key Laboratory of ASIC and System, School of Microelectronics, Fudan University, Shanghai, China.*
[11]*Institute of Nanoelectronics and Quantum Computing, Fudan University, Shanghai 200433, China*
[12]*Shanghai Branch, Hefei National Laboratory, Shanghai 201315, China*

# Abstract

The pursuit of discovering new high-temperature superconductors that diverge from the copper-based paradigm[1-3] carries profound implications for elucidating mechanisms behind superconductivity and may also enable new applications[4-8]. Here, our investigation reveals that application of pressure effectively suppresses the spin and charge order in trilayer nickelate $La_4Ni_3O_{10-\delta}$ single crystals, leading to the emergence of superconductivity with a maximum critical temperature ($T_c$) of around 30 K at 69.0 GPa. The DC susceptibility measurements confirm a substantial diamagnetic response below $T_c$, indicating the presence of bulk superconductivity with a volume fraction exceeding 80%. In the normal state, we observe a "strange metal" behavior, characterized by a linear temperature-dependent resistance extending up to 300 K. Furthermore, the layer-dependent superconductivity observed hints at a unique interlayer coupling mechanism specific to nickelates, setting them apart from cuprates in this regard. Our findings provide crucial insights into the fundamental mechanisms underpinning superconductivity, while also introducing a new material platform to explore the intricate interplay between the spin/charge order, flat band structures, interlayer coupling, strange metal behavior and high-temperature superconductivity.

Cuprates are the first family of high-temperature (high-$T_c$) superconducting materials, characterized by layers of $CuO_2$ interleaved with charge reservoir layers[1,2]. Despite intensive research on various cuprates, the mechanism responsible for high-temperature superconductivity remains elusive[2,3]. Consequently, the pursuit of high-temperature superconductors that do not rely on copper oxides has become a focal point of intense experimental and theoretical exploration since the discovery of cuprates nearly four decades ago[4-8]. This is motivated by the belief that such materials may help to elucidate the enigmatic mechanisms governing high-temperature superconductivity while opening doors to new applications. Nickel, situated immediately to the left of copper on the Periodic Table, offers a playground for materials and chemistry designs aimed at replicating high-temperature unconventional superconductivity[8-12]. However, despite intensive efforts, achieving superconductivity in nickelates has proven to be a formidable challenge. In 2019, an intriguing development occurred when superconductivity was observed in "infinite-layer" nickelate thin films with $T_c$ = 5-20 K[13-16]. In these materials, $Ni^{1+}$ ($d^9$) forms square planar $NiO_2$ layers closely resembling $Cu^{2+}$ ($d^9$) in cuprates[13-16].

More recently, signatures of superconductivity have also been observed in the Ruddlesden-Popper (RP) bilayer perovskite $La_3Ni_2O_7$ under high pressure, achieving a $T_c$ of approximately 80 K above 14 GPa (ref. 17). Subsequent studies have observed zero resistance under improved hydrostatic pressure conditions facilitated by a liquid pressure-transmitting medium[18-20]. However, it has been suggested that the superconductivity in $La_3Ni_2O_7$ may be filamentary in nature, with a low superconducting volume fraction[19,20], underscoring the need for more in-depth research to fully understand the superconducting properties of this class of materials. Unlike "infinite-layer" nickelates and cuprates with a $d^9$ electron configuration, $La_3Ni_2O_7$ hosts a bilayer $NiO_2$ square structure featuring $Ni^{2.5+}$ ($d^{7.5}$) ions[17]. Furthermore, the $p$-orbital of apical oxygen, which connects adjacent $NiO_2$ layers, couples the two nearest-neighbor $3d_{z^2}$ orbitals, suggesting that interlayer coupling may also play a crucial role in $La_3Ni_2O_7$[17]. However, in contrast to cuprate and iron-based superconductors, where superconductivity typically arises from the suppression of static long-range magnetic order in their parent phases[2-4,21], infinite-layer and bilayer $La_3Ni_2O_7$ nickelates have shown either a lack of magnetic order or hinted at the presence of weak magnetism[13,17,22,23]. This raises a fundamental question about whether magnetism plays the same crucial role in making nickelates into high-temperature superconductors.

Of particular interest, it is well known that the $T_c$ in cuprates depends on the number of $CuO_2$ layers ($n$) in a non-monotonic way, reaching a maximum for $n = 3$ in most cases[24-26]. As a result, trilayer cuprates have the highest $T_c$ among all cuprates, reaching up to 135 K at ambient pressure and 164 K under high pressure for mercury-based compounds[27,28]. The mechanism for this layer dependence of superconductivity remains a subject of intense and ongoing debate[26-30]. This engenders an intriguing question: can trilayer nickelates exhibit superconductivity, and if so, how might it influence $T_c$.

Theoretical considerations have suggested that trilayer $La_4Ni_3O_8$, with its $Ni^{1.33+}$ ($d^{8.67}$) configuration, closely parallels the cuprates with $Cu^{2+}/Cu^{3+}$ configurations and thus stands as an ideal candidate for a high-$T_c$ superconductor[31,32]. However, experimental investigations have thus far failed to observe superconductivity in trilayer $La_4Ni_3O_8$, both under ambient conditions and at high pressure[11,33].

In contrast to many nickelates that display insulating behavior, including trilayer $La_4Ni_3O_8$, the trilayer RP compound $La_4Ni_3O_{10}$ stands out as a rare oxide compound that maintains its metallic character even at low temperatures under ambient pressure[34,35]. Trilayer nickelate $La_4Ni_3O_{10}$ exhibits a static incommensurate magnetic order accompanied by a charge order[36]. The nominal valence state of trilayer $La_4Ni_3O_{10}$ is $Ni^{2.67+}$ ($d^{7.33}$), which is different from the $d^9$ state typically observed in infinite-layer nickelates and cuprates or the $d^{7.5}$ state found in bilayer $La_3Ni_2O_7$. Nevertheless, the band structure of $La_4Ni_3O_{10}$ unveils intriguing features: the $d_{x^2-y^2}$ hole band bears a striking resemblance to the behavior observed in hole-doped cuprates, while the $d_{z^2}$ band is rather flat and exhibits a 20 meV density wave-like gap opening associated with the spin order transition[35,36], reminiscent of phenomena observed in iron-based superconductors[21,37]. This distinctive combination of characteristics, coupled with its trilayer structure, positions $La_4Ni_3O_{10}$ as an ideal platform for the exploration of the interplay between magnetism, interlayer coupling, and potential superconductivity. However, the investigations of $La_4Ni_3O_{10}$ were significantly hampered by the scarcity of high-quality single crystals, which necessitated their growth under a high oxygen pressure atmosphere[34].

In this paper, we report detailed measurements of $La_4Ni_3O_{10-\delta}$ single crystals under both ambient conditions and high pressures, reaching up to 70 GPa. The high-quality $La_4Ni_3O_{10-\delta}$ single crystals were grown using a high-pressure vertical optical-image floating-zone furnace. Our X-ray diffraction (XRD) experiments conducted on powdered $La_4Ni_3O_{10-\delta}$ single crystals at ambient conditions confirm the presence of a pure phase of trilayer $La_4Ni_3O_{10-\delta}$ (Extended Data Fig. 1). Further, we conducted single-crystal neutron-diffraction and single-crystal XRD measurements on $La_4Ni_3O_{10-\delta}$ (Extended Data Fig. 7a, Supplementary Table 1). The combined single-crystal structural refinement analysis using least square fitting by incorporating data from both neutron diffraction and XRD, identified the composition as $La_4Ni_3O_{9.96(4)}$, pointing to a minor oxygen deficiency. Additionally, the refined crystal structure aligns with the $P2_1/a$ space group with $Z=2$ (Fig. 1e), consistent with previous reports[34].

Figure 1b displays atomic-resolution high-angle annular dark-field (HAADF) images, revealing the detailed positions of atoms within the trilayer structure of $La_4Ni_3O_{10-\delta}$. We also employed integrated differential phase contrast (iDPC) imaging techniques to accurately visualize the lighter oxygen atoms (Extended Data Fig. 6). The simultaneous acquisition of HAADF and iDPC images, seamlessly aligns with our neutron diffraction and XRD refinements, reinforcing the accuracy of our structural analysis. The HAADF images in large area overviews further confirm the phase purity and exceptional quality of our crystals (Extended Data Fig. 6).

To assess the influence of pressure on the crystal structure, we conducted synchrotron-based XRD measurements on powdered $La_4Ni_3O_{10-\delta}$ single crystals (Extended Data Fig. 3). It is shown that the diffraction peaks within the range of $12.5°<2\theta<14°$ exhibit a noticeable evolution with increasing pressure, indicative of a structural phase transition. The refinement analyses reveal a subtle structural transition associated with the tilting of the $NiO_6$ octahedra, shifting from monoclinic $P2_1/a$ to tetragonal $I4/mmm$ at pressures exceeding approximately 13-15 GPa (Extended Data Fig. 3). Concurrently, the Ni–O–Ni angle between adjacent $NiO_2$ layers undergoes a change from around 165.6(6)° to 180° during this phase transition (Fig. 1g), potentially enhancing interlayer coupling. The lattice constants and cell volume also exhibited progressive decrease under pressure, with an anomaly observed near the structural phase transition (Fig. 1c, 1d), which is consistent with a recent independent measurement on polycrystalline samples[38]. To ascertain if such a phase transition also occurs in single crystal samples, single-crystal XRD measurements were conducted, with the structural phase transition being further confirmed by single-crystal refinements (Extended Data Fig. 7b, Supplementary Table 2).

To further characterize the material, magnetic susceptibility measurements were conducted, revealing a distinct kink in the data at $T_N \approx 136$ K (Fig. 2a), suggesting the emergence of the static spin and charge order, as previously revealed through neutron diffraction experiments[36]. This phase transition was corroborated by heat capacity measurements, which exhibited a pronounced peak at a similar temperature (Fig. 2c). These results are consistent with the material's monoclinic $P2_1/a$ structure at ambient pressure[34,39].

The electrical resistance $R(T)$ of $La_4Ni_3O_{10-\delta}$ single crystals under various pressure conditions is presented in Fig. 2b and Fig. 3a-e. At ambient pressure, $La_4Ni_3O_{10-\delta}$ displays a characteristic metallic behavior, with resistivity exhibiting a decrease as the temperature descends below 300 K. A distinctive, steplike kink in the resistivity curve manifests at the spin/charge ordering temperature $T_N$ (Fig. 2b). It's worth emphasizing that the spin/charge order phase transition observed in our measurements displays an exceptional sharpness, which is notable in comparison to previous studies. This indicates the high quality of our $La_4Ni_3O_{10-\delta}$ single crystals, setting the stage for precise investigations into its physical properties under pressure.

When external pressure is exerted on the piston-cylinder cell, the characteristic kink related to the spin/charge ordering in the resistivity curve is rapidly suppressed (Fig. 3a), which is consistent with previous measurements on a powder sample below 1.28

GPa[40].

Resistances above 2.5 GPa were measured in a Be-Cu alloy diamond anvil cell (DAC). To ensure the best hydrostatic condition inside the DAC, we use helium as the pressure-transmitting medium. The resistance initially exhibits weak insulating behavior, accompanied by a peak anomaly associated with the spin order transition at 3.0 GPa (Fig. 3b). As pressure increases, metallic behavior is restored and the peak anomaly is progressively suppressed, and eventually a sharp drop in resistance below a critical temperature $T_c$ of 4.5 K is observed at 15.5 GPa (see Fig. 3b). With further increments in pressure, $T_c$ continues to rise, ultimately reaching a point where zero resistance is observed at pressures exceeding 43.0 GPa (Fig. 3b-d). This compellingly signifies the emergence of superconductivity, with the onset superconducting transition temperature ranging from 4.5 K to around 30 K.

We also conducted temperature-dependent resistance measurements in a DAC using KBr as the pressure-transmitting medium. Across the pressure range spanning from 2.2 to 24.6 GPa, a moderate decrease of resistance below a critical temperature of 2-20 K is observed (Fig. 3e). This together with the weak upturn in resistance in the normal state, indicates a limited superconducting volume, likely due to the filamentary nature of the superconductivity within this pressure range. As the pressure is raised further, a significant sharp reduction in resistance becomes evident at approximately 20-30 K, observed at pressures of 38.0 GPa and beyond (Fig. 3e) and the onset $T_c$ reaches 30.1 K at 69.0 GPa (Fig. 3e, Extended Data Fig. 5). However, small residue resistance below $T_c$ was observed in this configuration, likely attributed to the less hydrostatic conditions resulting from the pressure-transmitting medium KBr.

To provide additional confirmation of the pressure-induced superconductivity, we conducted ultrasensitive DC magnetic susceptibility measurements under high pressures within a custom-built miniature Be-Cu alloy DAC, using nitrogen as the pressure-transmitting medium to provide a hydrostatic pressure environment. A distinctive diamagnetic response at $T_c$ is clearly evident in the zero-field-cooled (ZFC) curve at 40.0 GPa, and as pressure increases, $T_c$ also rises, consistent with the resistance measurements, further confirming the emergence of superconductivity (Fig. 3f-i and Extended Data Fig. 8). We have estimated the maximum superconducting volume fraction to be around 86% across the pressures applied, suggesting the development of bulk superconductivity (Extended Data Fig. 9). Conversely, below 30 GPa, no clear superconducting diamagnetic responses were detected, suggesting that the superconducting volume fraction is relatively limited at low pressures (Extended Data Fig. 10). The difference of susceptibilities between ZFC and field-cooled (FC) curves in the normal states is due to the magnetic background of the pressure cell with residual magnetic impurities (Supplementary Fig. 5).

Interestingly, concomitant with the emergence of zero resistance at 43.0 GPa, the normal state resistance follows a linear temperature dependence up to 300 K (Fig. 3b). This behavior is a hallmark of the so-called "strange metal" state, a characteristic phenomenon observed in optimal doped cuprates, certain iron-based materials and nickelate superconductors [2,17,18,41-45], implying the existence of strong correlations and underscores the unconventional nature of superconductivity. Similar strange metal behavior was also confirmed in the measurements conducted within the KBr DAC above 38.0 GPa (Fig. 3e).

Figs. 4a-d display the temperature-dependent magnetoresistance measured in magnetic fields perpendicular to the *ab* plane under various pressures. As the magnetic field is increased, superconductivity is progressively suppressed, providing further confirmation that the transition in resistance is due to the onset of superconductivity. We use the 90% resistance transition to the normal state near $T_c$ and fit it to the Ginzburg-Landau form $H_{c2}(T) = H_{c2}(0)[(1-t^2)/(1+t^2)]$, where t = $T/T_c$. This analysis yields an estimation of the upper critical field, with values reaching 44 T at 63.0 GPa when utilizing the helium DAC, and 48 T at 69.0 GPa with the KBr DAC (Fig. 4e, Extended Data Fig. 5). These values of upper critical field exceed those observed in infinite-layer nickelates[13] but fall below $La_3Ni_2O_7$[17].

Figure 1a summarizes the pressure dependent spin/charge order and superconductivity phase diagram in trilayer $La_4Ni_3O_{10}$. This phase diagram bears some analogy to those found in cuprate and iron-based superconductors[2,21], where high-temperature superconductivity arises upon suppression of a static magnetic order. From the electronic structure perspective, it suggests that the $d_{z^2}$ band could play an important role in shaping the pressure-dependent phase diagram of $La_4Ni_3O_{10}$. The $d_{z^2}$ band is notably flat and displays a 20 meV spin density wave-like gap[35]. This unique electronic structure feature, coupled with its propensity to interact with the *p*-orbital of apical oxygen, renders the $d_{z^2}$ band highly susceptible to external pressure, with the potential to alter nesting conditions and influence the spin/charge density wave order.

Therefore, the concurrent emergence of bulk superconductivity and strange metal behavior could be attributed to the pressure effect that suppresses the density wave gap, brings the flat $d_{z^2}$ band into proximity with the Fermi surface, consequently inducing strong correlations and fostering the emergence of the "strange metal" behavior. Simultaneously, the closing of the density wave gap dampens the static spin-density-wave order and promotes dynamic spin fluctuations, paving the way for superconductivity to emerge, where the hole Fermi surface associated with the $d_{x^2-y^2}$ orbital may also come into play.

If this analysis were extended to the bilayer system, it suggests that the absence of (or weak) static magnetic order in $La_3Ni_2O_7$ may be attributed to the fact that the $d_{z^2}$ band lies considerably further below the Fermi level compared to $La_4Ni_3O_{10}$[35,46]. However, even the flat bands located away from the Fermi level have the potential to generate high-energy spin fluctuations, which could exert a noteworthy influence on the phase diagram or serve as a mediator of electron pairing[47]. Further investigations in this direction are warranted to fully elucidate the role of magnetism in nickelate superconductors.

Furthermore, it's important to consider the impact of the interlayer coupling, a factor that could potentially promote superconductivity and has been intensively discussed in multilayer cuprates and bilayer $La_3Ni_2O_7$[2,17,29,30,48]. The observation of pressure induced superconductivity accompanied by a structural transition from monoclinic to tetragonal phases also suggests a potential significant role of interlayer coupling in trilayer $La_4Ni_3O_{10-\delta}$. However, unlike cuprate superconductors where the highest $T_c$ is achieved in trilayer systems, in the case of trilayer $La_4Ni_3O_{10-\delta}$, the $T_c$ is lower than that of bilayer $La_3Ni_2O_7$. This discrepancy suggests the presence of distinct interlayer interaction mechanisms between these two systems. Additional investigations are needed to elucidate these coupling mechanisms, particularly focusing on differences in carrier concentrations and magnetic structures between the inner and outer $NiO_2$ layers, as well as the interlayer coupling between the two outer $NiO_2$ layers. These factors are crucial for understanding the evolution of $T_c$ in multilayer superconductors[2,29,30,48].

In summary, we present evidences of bulk superconductivity in trilayer nickelate $La_4Ni_3O_{10-\delta}$ single crystals under pressure. Our experiments also unveil intriguing "strange metal" behavior in the normal state, characterized by linear temperature dependence of resistance up to 300 K, which may be linked to the enhanced spin fluctuations and strong correlations induced by the flat $d_{z^2}$ band positioned near the Fermi level. Furthermore, the layer-dependent $T_c$ in nickelates is distinct from that observed in cuprates, suggesting unique interlayer coupling and charge transfer mechanisms specific to nickelates. Further research is required to fully understand the precise role of interlayer coupling in the pairing, especially considering the differences in carrier concentrations and magnetism between the inner and outer $NiO_2$ layers, as well as the interlayer coupling between the two outer $NiO_2$ layers, which are absent in the bilayer system. Additionally, a comprehensive exploration of the role of the $d_{x^2-y^2}$ orbital and pairing symmetry is necessary for a complete understanding. In essence, our findings establish a promising new material platform, inviting deeper exploration into the intricate interplay between spin/charge order, flat band structure, interlayer coupling, strange metal behavior and high-temperature superconductivity. This avenue of research holds great potential for uncovering novel phenomena and advancing our understanding of high-temperature superconductors.

**Methods**

**Growth of $La_4Ni_3O_{10-\delta}$ single crystals**

The precursor powder for the $La_4Ni_3O_{10-\delta}$ compound was prepared using the conventional solid-state reaction method. Chemically stoichiometric raw materials, $La_2O_3$ and NiO (Aladdin, 99.99%), were meticulously ground and mixed using a Vibratory Micro Mill (FRITSCH PULVERISETTE 0). An additional 0.5% of NiO was included to compensate for potential NiO volatilization during the crystal growth process. The resulting mixture underwent calcination at 1373 K for 24 hours, with two repeated calcination cycles to ensure complete and homogeneous reactions.

Subsequently, the resulting precursor material was pressed into a cylindrical rod, approximately 13 cm in length and 6 mm in diameter, using a hydrostatic pressure of 300 MPa. The shaped rod then underwent once sintering at 1673 K for 12 hours in air. Single crystals were grown using a vertical optical-image floating zone furnace (Model HKZ, SciDre GmbH, Dresden) at Fudan University. During the crystal growth process, we carefully maintained an oxygen pressure within the range of 18-22 bar, and employed a 5 kW Xenon arc lamp as the light source. The rod was rapidly traversed through the growth zone at a speed of 15 mm/h to enhance the density, after which a growth rate of 3 mm/h was maintained.

**STEM measurements**

The STEM experiments were performed on a double aberration corrected S/TEM (Themis Z, Thermo Fisher Scientific) operated at 300 kV. For STEM-HAADF imaging, the probe semi-convergent angle is 21.4 mrad, and the HAADF collection angle is from 79 to 200 mrad. The atomic resolution EDX imaging was recorded with a Super-X detector.

**Measurements under high pressure**

We conducted electrical resistance measurements on $La_4Ni_3O_{10-\delta}$ single crystals under pressure in a physical property measurement system (PPMS) by Quantum Design. The temperature range covered was from 1.8 to 300 K, and magnetic fields up to 9 T were applied. The electrical resistivity measurements at low pressures (below 2.5 GPa) were performed in a piston cylinder cell using the standard four-probe method (Fig. 3a). The pressure-transmitting medium employed in this setup was liquid Daphne 7373. The pressure was calibrated from the superconducting transition temperature of Pb. The electrical resistance measurements at high pressures (above 2.2 GPa) were performed in a diamond anvil cell (DAC) with 200-300 μm culets using the standard van der Pauw four-probe method (Fig. 3b-d, Fig. 4, Extended Data Fig. 4). The sample chamber was constructed using a mixture of cubic boron nitride and epoxy, with a diameter ranging from 140 to 280 μm. The single crystal was loaded under helium as the pressure-transmitting medium. Alternately, KBr powders were also used as a different pressure-transmitting medium (Fig. 3e, Extended Data Fig. 5). Pressure calibration was carried out using ruby fluorescence peak shift at room temperature.

Ultrasensitive magnetic susceptibility measurements under pressure were conducted using a custom-built beryllium-copper alloy mini-DAC equipped with a rhenium gasket, employing a design similar to refs. 49 and 50. The mini-DAC has dimensions of approximately 8.5 mm in diameter and 30 mm in length. The measurements were performed using a Magnetic Property Measurement System (MPMS3, Quantum

Design). The DAC includes a pair of diamond anvils with a diameter of 300 μm and a sample chamber with a diameter of 210 μm. The sample chamber was filled with a single crystal of $La_4Ni_3O_{10-\delta}$ and liquid nitrogen as the pressure transmitting medium to provide a hydrostatic pressure environment. The samples had a diameter of approximately 150-200 μm and a thickness of roughly 25 μm.

The single-crystal neutron-diffraction experiments were performed using the HB-3A four-circle single-crystal neutron diffractometer at the High Flux Isotope Reactor at Oak Ridge National Laboratory. A Si (220) monochromatic neutron beam with a wavelength of 1.533 Å was employed for the measurements. *In-situ* lab-based high-pressure XRD measurements on powder and single crystals were carried out using a Bruker D8 Venture diffractometer, utilizing Mo Kα radiation ($\lambda = 0.7107$ Å) in a diamond anvil cell with 300-μm-diameter culets and stainless steel gaskets (Extended Data Fig. 2, Extended Data Fig. 7, Supplementary Fig. 9, Supplementary Table 1 and 2). A methanol-ethanol-water mixture in the ratio of 16:3:1 was used as the pressure-transmitting medium for these measurements. For the powder measurements, we captured two-dimensional patterns of powder diffraction rings, which were subsequently processed to yield one-dimensional profiles (Extended Data Fig. 2). We included a small ruby ball within the chamber to facilitate pressure calibration by monitoring the ruby's fluorescence peak shift. *In-situ* synchrotron-based high-pressure XRD measurements were performed at the beamline 15U1 at Shanghai Synchrotron Radiation Facility, China, utilizing an X-ray beam with a wavelength of 0.6199 Å (Extended Data Fig. 1b, Extended Data Fig. 3, Supplementary Table 3). A symmetric DAC with anvil culet sizes of 300 μm, and rhenium gaskets were used. Helium was used as the pressure-transmitting medium to maintain optimal hydrostatic pressure conditions, with pressure calibration again based on the shift in the fluorescence peak of a ruby indicator.

**Calculation of superconducting volume fraction**

The susceptibility $\chi_m$ of a superconductor can be decomposed into:

$$\chi_m = (1-f)\chi_p + f\chi_d, \quad (1)$$

where $f$ is the superconducting volume fraction, $\chi_p$ denotes the paramagnetic susceptibility, and $\chi_d$ denotes the diamagnetic susceptibility.

In SI units, $\chi_d = -1$, and $\chi_p \to 0$, eq. (1) can be then expressed as

$$f = |\chi_m|, \quad (2)$$

Upon converting between unit systems, the susceptibility equation becomes:

$$\chi_{SI} = 4\pi\chi_{CGS} = 4\pi M/H = 4\pi\Sigma\mu_i/\Sigma V_i = (4\pi\mu m_a)/(HVN_a m), \quad (3)$$

where $\mu \approx -1.69\times10^{-6}$ emu represents the measured magnetic moment at 5 K after subtracting the constant background signal at onset of $T_c$ for sample S6 at 20 Oe and 50 GPa. $m \sim 3.16\ \mu g$ is the sample mass which was calculated based on the refined density $\rho = 7.16$ g/cm$^3$ and the sample volume at ambient condition. $m_a$ is the molar mass, $N_a$ is the Avogadro constant, $V = 342.837$ Å$^3$ is the cell volume for a $I4/mmm$ primitive cell with $Z = 2$ at 50 GPa (Fig. 1d, Supplementary Table 3).

The demagnetizing factor $N$ of a finite cylinder with axial magnetic field can be approximated using[51]:

$$N^{-1} \approx 1 + 1.6\ (C/A), \quad (4)$$

where $A$ is the diameter and $C$ is the thickness of the sample. For sample S6, $A = 160\ \mu m$ and $C = 22\ \mu m$ at ambient condition, yielding a demagnetizing factor of $N = 0.82$. This factor remains largely unchanged under hydrostatic pressure due to the proportional contraction of the lattice dimensions of the crystal structure (Supplementary Table 3). The susceptibility $\chi$ can be calculated as:

$$N = 1/\chi_0 - 1/\chi, \quad (5)$$

where $\chi_0$ is the measured value and $\chi$ is the adjusted susceptibility for the sample geometry. Implementing the above procedure, $\chi$ calculates to approximately -0.86 at 5 K for S6, indicating a superconducting volume fraction of 86% as shown in Extended Data Fig. 9. This substantiates pressure-induced bulk superconductivity in $La_4Ni_3O_{10-\delta}$.

**Extended analysis and discussion**

The upper critical fields of nickelates are comparable with those observed in hole-doped cuprate and iron-pnictide superconductors exhibiting similar $T_c$ values[52-54], but surpass those of simple conventional superconductors[49,55]. Furthermore, our estimation of the in-plane superconducting coherence length of $La_4Ni_3O_{10-\delta}$ approximates to 26 Å at 69.0 GPa, close to values of 122 iron pnictide and $La_{2-x}Sr_xCuO_4$ (refs. 52-54). The high upper critical fields and the correspondingly shorter superconducting coherence lengths point to tightly bound Cooper pairs, hinting at an unconventional pairing mechanism possibly driven by spin fluctuations and strong electronic correlations. Indeed, theoretical calculations on both infinite-layered nickelates and bilayer nickelates suggest that electron-phonon coupling alone is insufficient to account for the observed high critical temperatures[56,57]. We anticipate that the scenario for $La_4Ni_3O_{10}$ is similar, although specific calculations for this material have not yet been published in the literature.

Recently, a growing number of theoretical preprints on $La_4Ni_3O_{10}$ have appeared, examining its intricate electronic structure, the unique layer-dependent magnetism of its trilayer configuration, and its superconductivity. These analyses underscore the critical impact of electronic correlations, spin fluctuations, and interlayer coupling on the material's "strange metal" behavior and its unconventional superconducting pairing mechanism, with an emphasis on the $s_{\pm}$-wave pairing symmetry[58-64]. It has been suggested that the magnetic structures in $La_4Ni_3O_{10}$ vary substantially between the inner and outer $NiO_2$ layers, where the inner layer is non-magnetic and the outer layers are antiferromagnetic[61-63]. There are also likely variations in carrier concentrations and electron correlations across these layers[58-64]. These distinctions in magnetic structure and electron dynamics potentially introduce unique phenomenon not present in bilayer and infinite-layer nickelates, warranting further exploration.

While preparing this manuscript, we became aware of three related preprints that reported resistance measurements on $La_4Ni_3O_{10}$ polycrystalline samples[65-67]. References 66 and 67 describe a reduction in resistance—ranging from 4-7% below the 15-20 K temperature range—under pressures exceeding roughly 30 GPa, while ref. 65 does not observe this phenomenon below 50 GPa.

[†] These authors contribute equally to this work.
[*] Correspondence and requests for materials should be addressed to J.Z. (zhaoj@fudan.edu.cn) or J.G.G. (jgguo@iphy.ac.cn) or Q.S.Z.(zengqs@hpstar.ac.cn)

**Acknowledgments** This work was supported by the Key Program of the National Natural Science Foundation of China (Grant No. 12234006), the National Key R&D Program of China (Grant No. 2022YFA1403202), the Innovation Program for Quantum Science and Technology (Grant No. 2024ZD0300103), the Beijing Natural Science

Foundation (Grant No. Z200005), and the Shanghai Municipal Science and Technology Major Project (Grant No. 2019SHZDZX01). Y.H.Z. was supported by the Youth Foundation of the National Natural Science Foundation of China (Grant No. 12304173). H.L.W. was supported by the Youth Foundation of the National Natural Science Foundation of China (Grant No. 12204108). B.Y.P. was supported by the Natural Science Foundation of Shandong Province (Grant No. ZR2020YQ03). D.P and Q.S.Z. acknowledge the financial support from the Shanghai Science and Technology Committee (No. 22JC1410300) and Shanghai Key Laboratory of Material Frontiers Research in Extreme Environments (No. 22dz2260800). A portion of this work was carried out at the Synergetic Extreme Condition User Facility (SECUF). A portion of this research used resources at the High Flux Isotope Reactor, a DOE Office of Science User Facility operated by the Oak Ridge National Laboratory. A portion of this research used resources at the beamline 15U1 of Shanghai synchrotron radiation facility.

**Author contributions** J.Z. planed the project. Y.H.Z., E.K.Z. and L.X.C. synthesized the single-crystal samples. Y.H.Z., F.Y.L., E.K.Z., L.X.C., H.L.W., Y.Q.G. and Y.M.G. performed the thermodynamic and transport measurements at ambient conditions. E.K.Z., B.Y.P., X.C., D.P., W.B.W. and J.J.W. performed the resistance measurements under pressure with the support of J.G.G., Q.S.Z., L.J. and J. Z.. H.F.R., T.P.Y., X.L.C. and Y.S. assisted the resistance measurement under pressure. D.P. and Z.F.X. conducted the susceptibility measurements under pressure with the support of Q.S.Z.. N.N.L., Y.H.Z., D.P. and F.J.L. performed the synchrotron XRD measurements and analysis with the support of W.G.Y., Q.S.Z.. J.Y.H. and Y.H.Z. conducted the S/TEM characterization with the support of C.L.Z.. D.H.J., Y.H.Z., and Y.Q.H. performed the in-house powder and single crystal diffraction measurements under pressure and data analysis with the support of H.Y.G., H.B.C., and J.Z.. Y.Q.H. performed the neutron single crystal diffraction measurements and data analysis. J. Z., Y.H.Z. and F.Y.L. analyzed the data. J.Z., B.Y.P. and Y.H.Z. wrote the paper. All authors provided comments on the paper.

**Competing interests** The authors declare no competing interests.

**Data availability** Source data are provided with this paper.

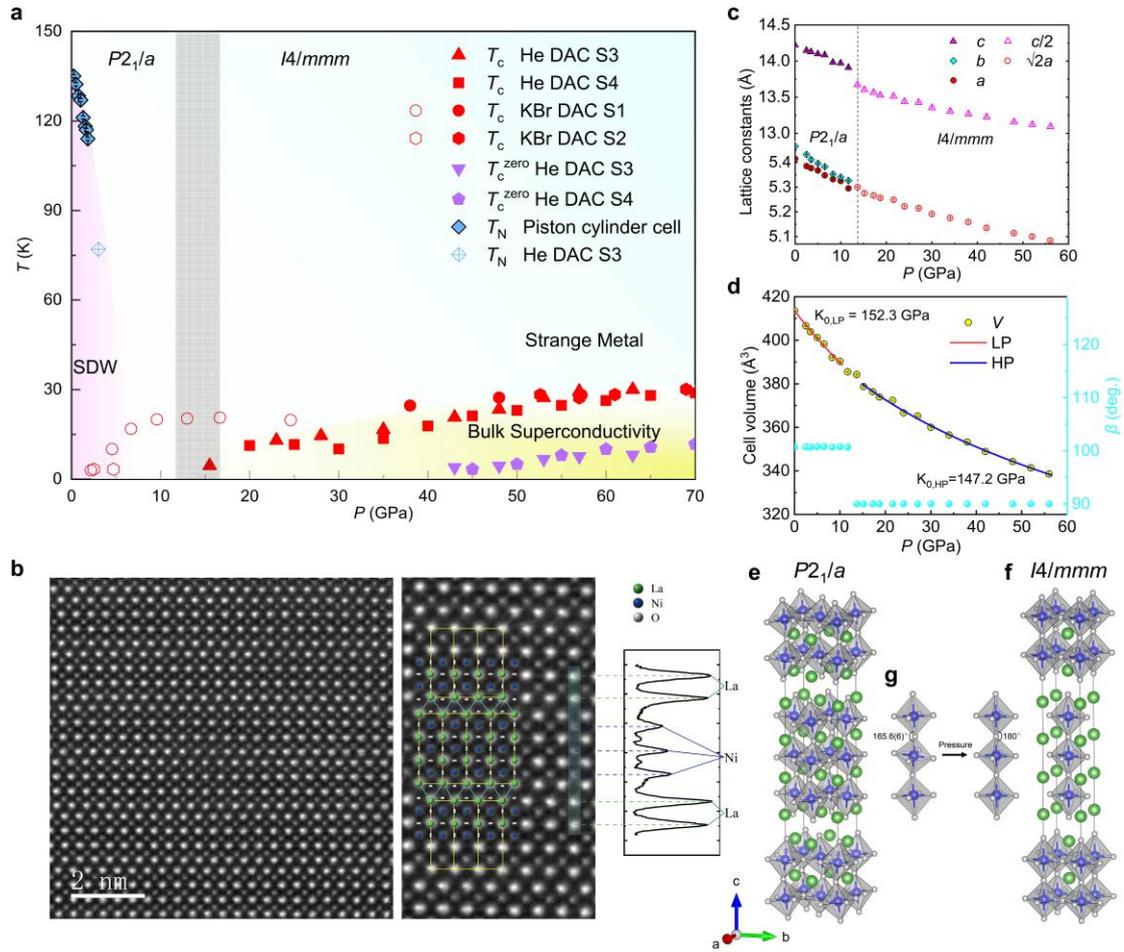

**Figure 1 │** Pressure-dependent lattice structure and phase diagram of $La_4Ni_3O_{10-\delta}$. **a,** Phase diagram of $La_4Ni_3O_{10-\delta}$ under pressure. The red solid triangles and squares represent $T_c$ (onset) of sample S3 and sample S4 in the helium DAC, respectively. The $T_c$ (onset) is defined as the temperature below which the resistance deviates from its linear dependence at high temperature. The red solid circles and hexagons represent the $T_c$ (onset) at pressures above 38 GPa of sample S1 and sample S2, respectively, where a pronounced sharp drop in resistance below $T_c$ is evident in the KBr DAC. The red open circles and hexagons denote the $T_c$ (onset) at pressures below 25 GPa in the KBr DAC, where a moderate decrease in resistance below $T_c$ is observed due to filamentary superconductivity (Fig. 3e). The blue diamonds denote the $T_N$ determined from resistance measurements in Figs. 3a and 3b, and Supplementary Fig. 7. Purple solid downward triangles and pentagons represent the $T_c$ offset $T_c^{zero}$ where resistance equals zero. Bulk superconductivity emerges above around 40 GPa, where substantial superconducting diamagnetic responses and zero resistance are observed. Shaded area highlights the region of the structural transition. **b**, Atomic-resolution high-angle annular dark-field images along the [110] direction at ambient pressure, featuring three layers of $NiO_2$ separated by LaO spacers. **c**, Lattice constants $a$, $b$, $c$ extracted from the synchrotron-based XRD (Extended Data Figure 3). **d,** Cell volume $V$ and the refined β angle of $P2_1/a$ and $I4/mmm$ lattices. Solid lines indicate the Birch-Murnaghan equation fit of cell volume as a function of pressure, with fitted bulk moduli $K_0$ of 152.3 GPa for the low-pressure phase and 147.2 GPa for the high-pressure phase. **e**, Crystal

structure of $La_4Ni_3O_{10-\delta}$ at ambient pressure. **f**, Crystal structure of $La_4Ni_3O_{10-\delta}$ at 34 GPa. **g**, Evolution of Ni-O-Ni angle between adjacent $NiO_2$ layers across the structural phase transition. Error bars, 1 s.d.

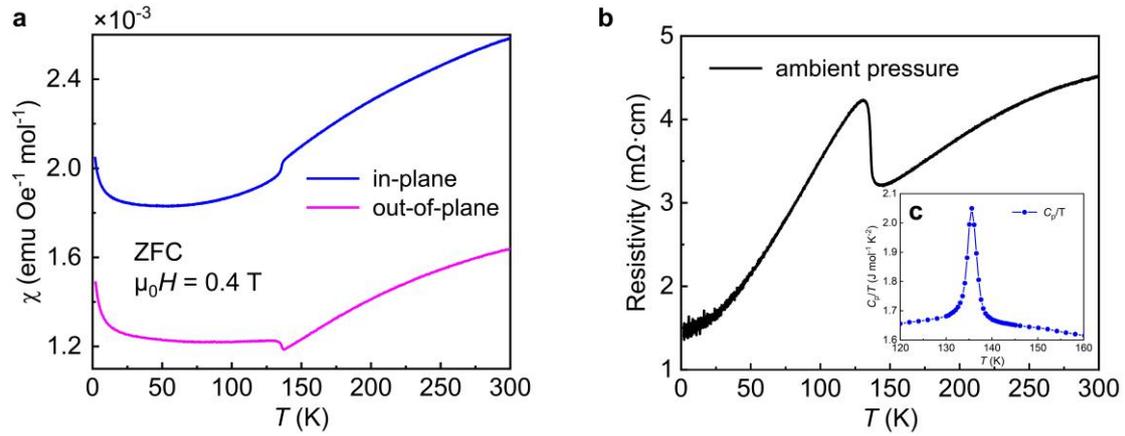

**Figure 2** | Magnetic susceptibility, resistivity and specific heat of $La_4Ni_3O_{10-\delta}$ single crystal at ambient pressure. **a,** Magnetic susceptibility of $La_4Ni_3O_{10-\delta}$ measured from 2 to 300 K with an applied field of 0.4 T, parallel and perpendicular to the *ab* plane. The SDW/CDW transition characterized by a kink in the $\chi(T)$ curve occurs at $T_N \approx 136$ K. **b,** Resistivity profile of $La_4Ni_3O_{10-\delta}$ in the *ab* plane at ambient pressure, utilizing a current of 100 μA. **c,** Specific heat of $La_4Ni_3O_{10-\delta}$ near $T_N$.

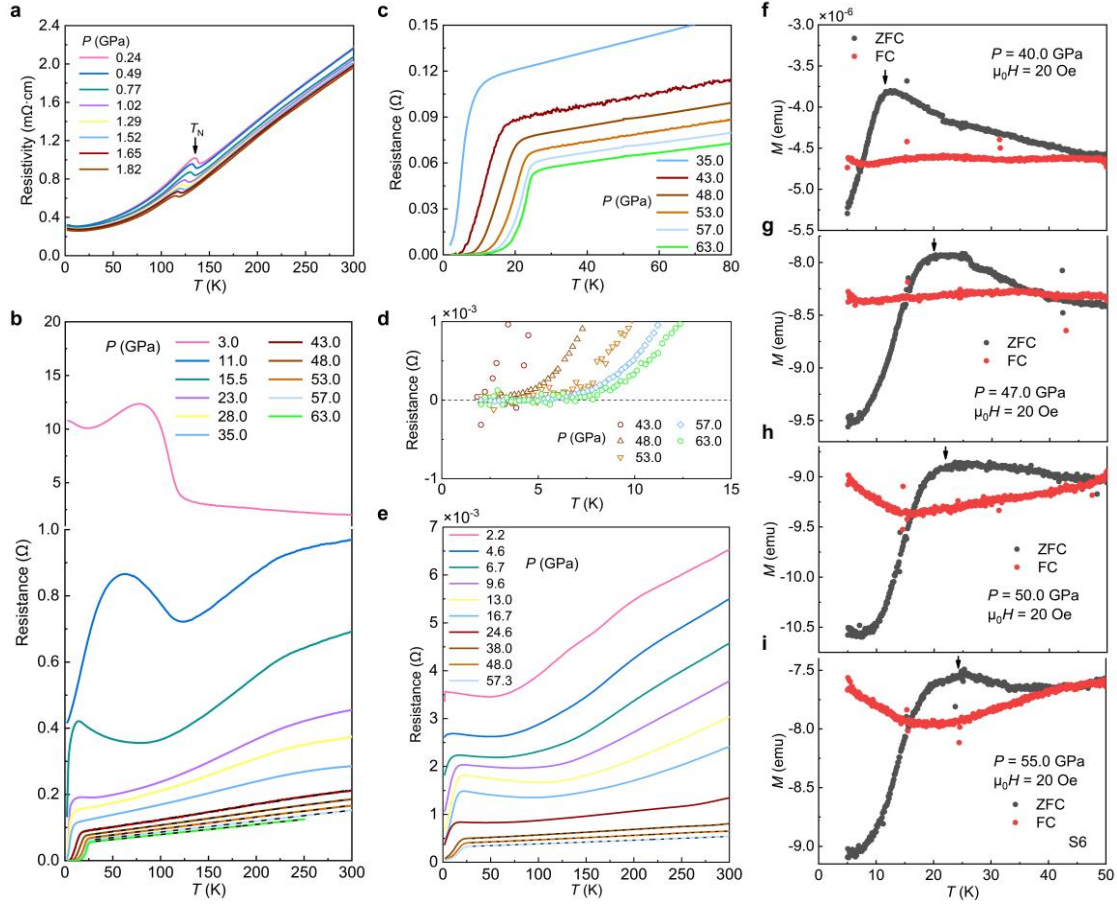

**Figure 3 | Temperature-dependendent resistances and DC susceptibilities of $La_4Ni_3O_{10-\delta}$ single crystals under various pressures. a,** Resistivity at pressures spanning 0.24 to 1.82 GPa in the piston cylinder cell, utilizing a current of 2 mA. The SDW/CDW transition at $T_N$ is progressively suppressed with increasing pressure. **b,** Resistances at pressures ranging from 3.0 to 63.0 GPa in the helium DAC for sample 3 (S3), utilizing a current of 500 μA. The diamond broke upon warming to around 250 K at 63.0 GPa. The black dashed lines depict the linear fit of the normal state resistances. **c,** Resistances in the helium DAC near the superconducting transition for S3. Zero resistances are observed above 43.0 GPa. **d,** An enlarged view of the resistance curve below $T_c$ within the helium DAC, providing a clear and evident demonstration of zero resistance for S3. **e,** Resistances at pressures ranging from 2.2 to 57.3 GPa in the KBr DAC for sample S1, utilizing a current of 100 μA. The black dashed lines depict the linear fit of the normal state resistances. **f,** Temperature dependent magnetization curves of $La_4Ni_3O_{10-\delta}$ under a magnetic field of 20 Oe applied perpendicular to the *ab* plane using the zero-field-cooled (ZFC) and field-cooled (FC) modes at 40.0 GPa for sample S6. A distinct superconducting diamagnetic response at $T_c$ is clearly observed in the ZFC curve. **g,** 47.0 GPa. **h,** 50.0 GPa. **i,** 55.0 GPa. The black arrows indicate the superconducting transition temperatures.

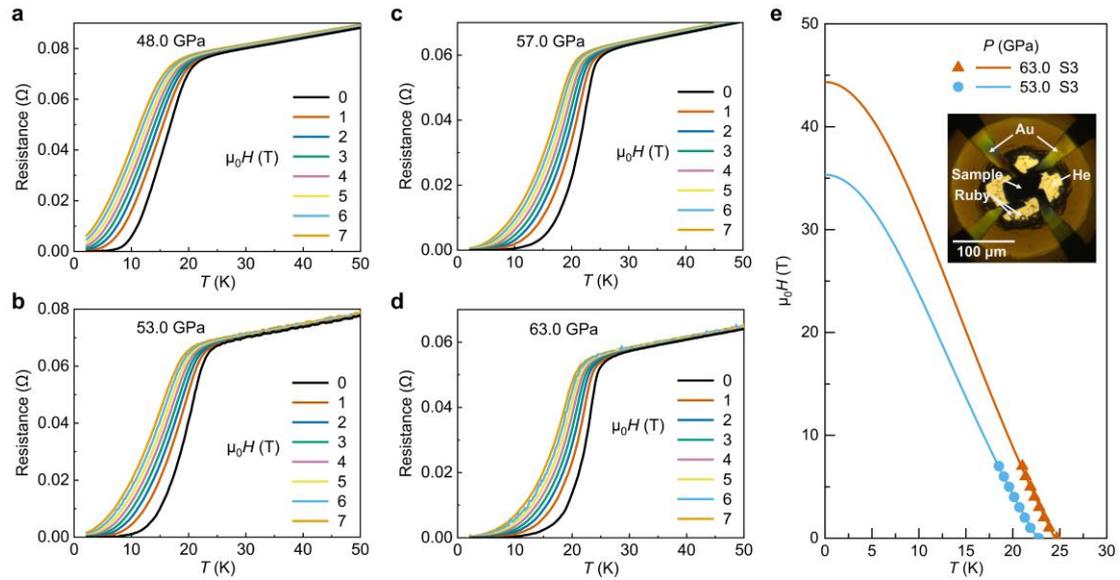

**Figure 4** | Magnetic field effects on the superconducting transition in $La_4Ni_3O_{10-\delta}$. **a,** Field dependences of electrical resistance at 48.0 GPa for sample S3. **b,** 53.0 GPa. **c,** 57.0 GPa. **d,** 63.0 GPa; the inset shows a photograph of the electrodes used for high-pressure resistance measurements. **e,** The Ginzburg–Landau fittings of the upper critical fields at 53.0 and 63.0 GPa. The magnetic fields are applied perpendicular to the *ab* plane.

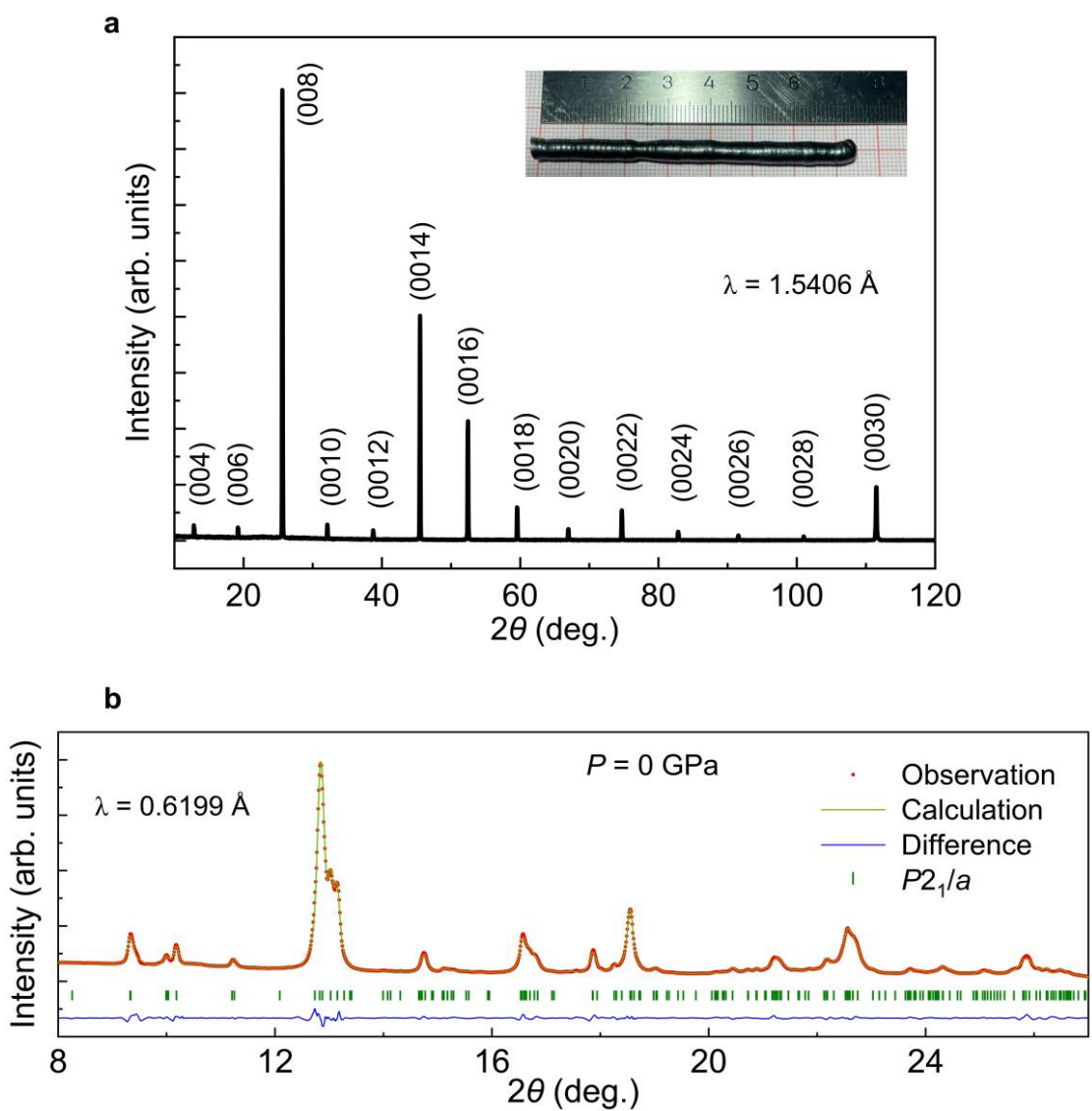

**Extended Data Fig. 1** | XRD measurements of $La_4Ni_3O_{10-\delta}$ single crystals and powdered single crystals. **a**, XRD measurements of a $La_4Ni_3O_{10-\delta}$ single crystal along the *ab* plane, revealing no detectable impurity phases. The measurements were carried out on a Bruker D8 Discover diffractometer, utilizing Cu Kα radiation. The inset displays a photograph of a $La_4Ni_3O_{10-\delta}$ single crystal, grown using the high-pressure floating-zone method. **b**, Rietveld refinement of synchrotron-based powder XRD pattern of $La_4Ni_3O_{10-\delta}$ at ambient pressure and room temperature. The olive solid lines and red circles represent the fitted and experimental data, respectively. The blue solid lines indicate the intensity difference between the data and calculations. The short vertical bars indicate the Bragg peak positions. The data can be well described with the space group $P2_1/a$. The data were collected using a powdered $La_4Ni_3O_{10-\delta}$ single crystal. The refinement parameters are summarized in Supplementary Table 3.

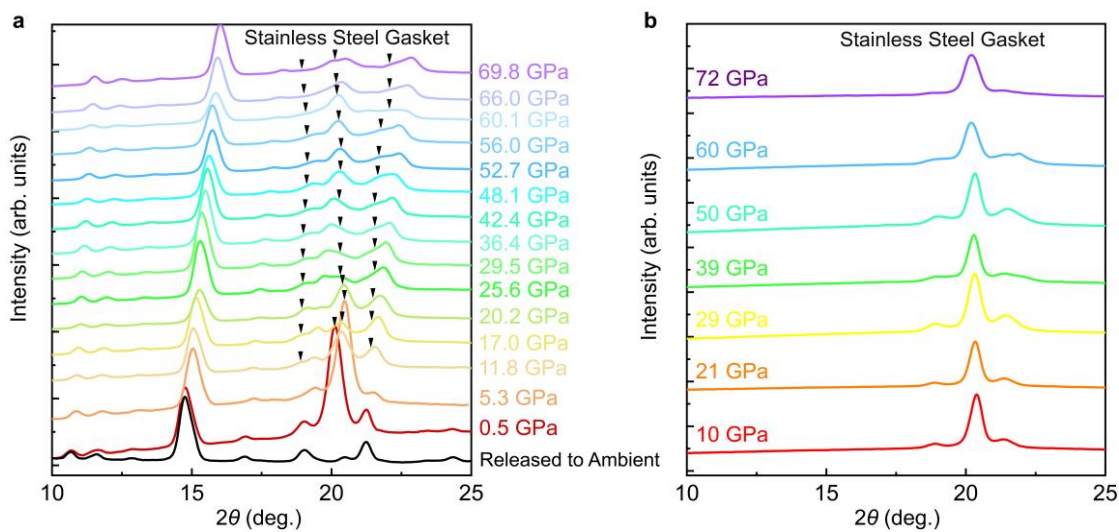

**Extended Data Fig. 2** | Lab-based XRD measurements of a powdered $La_4Ni_3O_{10-\delta}$ single crystal at various pressures at room temperature. **a**, XRD measurements at various indicated pressures. The black arrows indicate the intensities from stainless steel. Upon releasing pressure to ambient condition, the diffraction pattern (black solid line) remains essentially unaltered in comparison to the data obtained prior to applying pressure, indicating that the process is reversible and the sample is stable against pressure. The structural transition observed in synchrotron-based XRD was not resolved in this measurement due to limited resolution. **b**, Measurements of empty cell with stainless steel gasket at indicated pressures at room temperature.

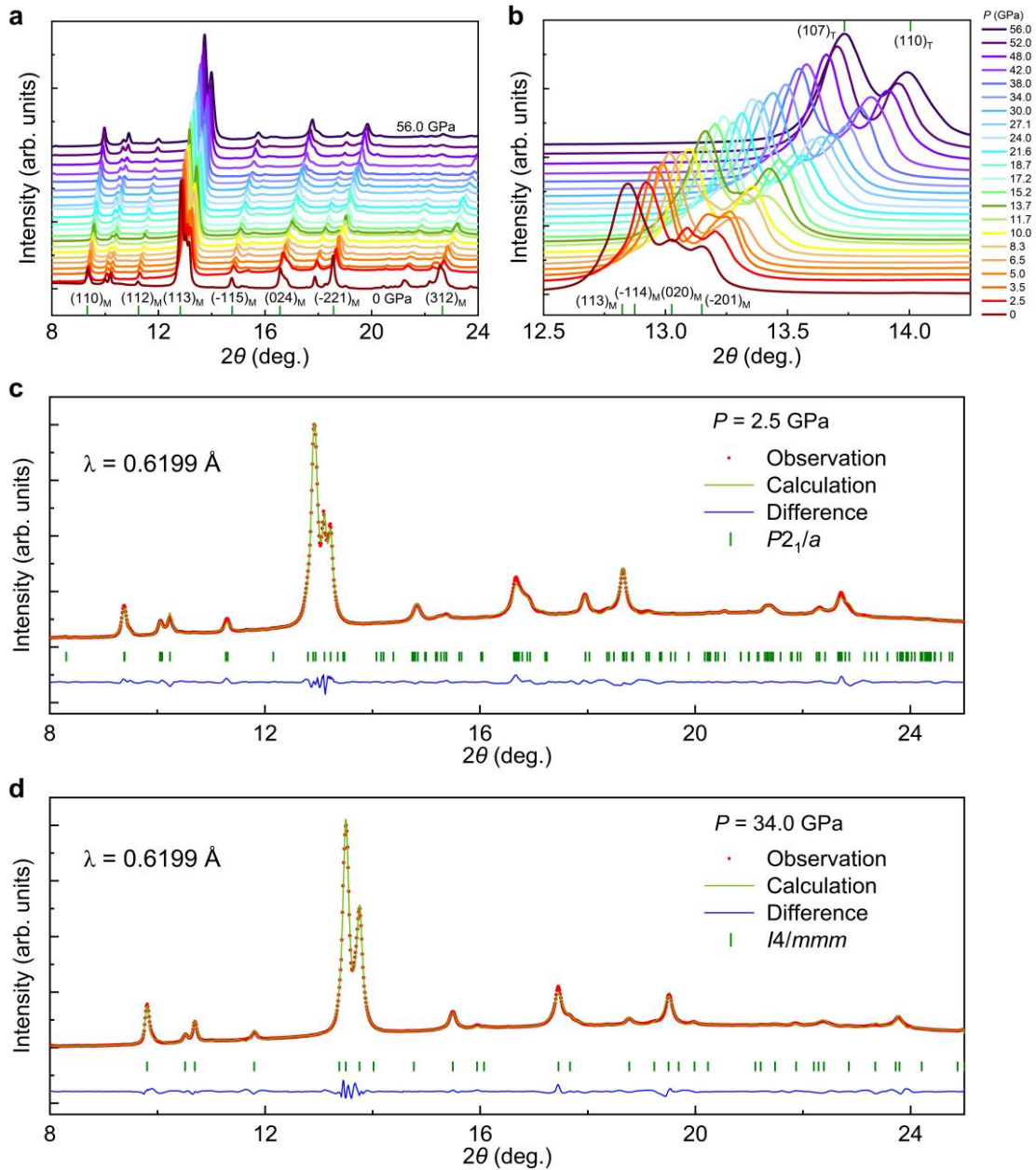

**Extended Data Fig. 3** | Synchrotron-based XRD measurements of powdered La$_4$Ni$_3$O$_{10-\delta}$ single crystals under various pressures, using a helium DAC at room temperature. **a,** XRD measurements under various pressures. **b**, An enlarge view of diffraction peaks within the range of 12.5°<2θ<14°, illustrating the merging of the monoclinic (0 2 0)$_M$ and (-2 0 1)$_M$ peaks into the tetragonal (1 1 0)$_T$ peak, a clear indication of the structural phase transition from monoclinic to tetragonal. **c**, Rietveld refinement of XRD pattern of La$_4$Ni$_3$O$_{10-\delta}$ at 2.5 GPa. **d**, Rietveld refinement of XRD pattern of La$_4$Ni$_3$O$_{10-\delta}$ at 34.0 GPa. The refinement parameters are summarized in Supplementary Table 3.

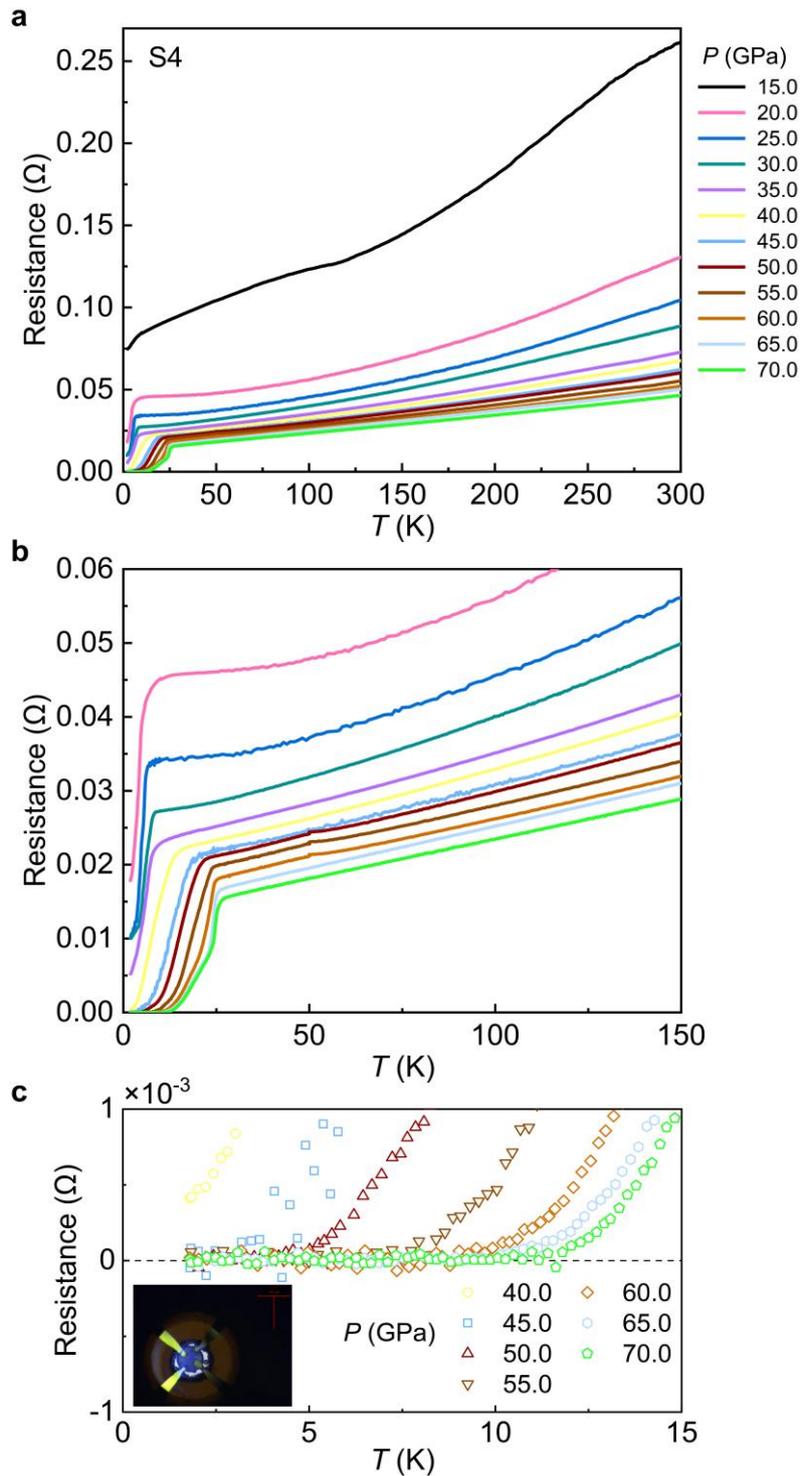

**Extended Data Fig. 4** | Temperature-dependent resistances of sample 4 (S4) in a helium DAC **a**, Temperature-dependent resistances of S4 under various pressures between 2 K and 300 K. **b**, Detailed resistance profile from 2 K to 150 K. Zero resistance below $T_c$ is observed above 45 GPa. **c**, An enlarged view of the resistance curve below $T_c$, providing a closer examination of the zero-resistance state. The inset is a photograph depicting the electrodes used for the measurement.

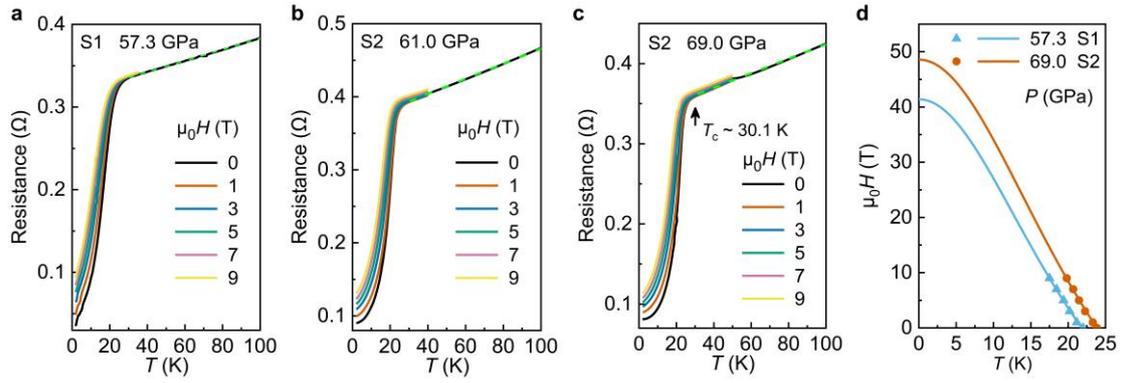

**Extended Data Fig. 5 | Magnetic field effects on the superconducting transition in $La_4Ni_3O_{10-\delta}$ in a KBr DAC. a**, Field dependences of electrical resistance at 57.3 GPa for sample S1. **b,** Field dependences of electrical resistance at 61.0 GPa for sample S2. **c,** Field dependences of electrical resistance at 69.0 GPa for sample S2. The green dashed lines depict the linear fit of the normal state resistances. **d,** The Ginzburg–Landau fittings of the upper critical fields at 57.3 (S1) and 69.0 GPa (S2). The magnetic fields are applied perpendicular to the *ab* plane.

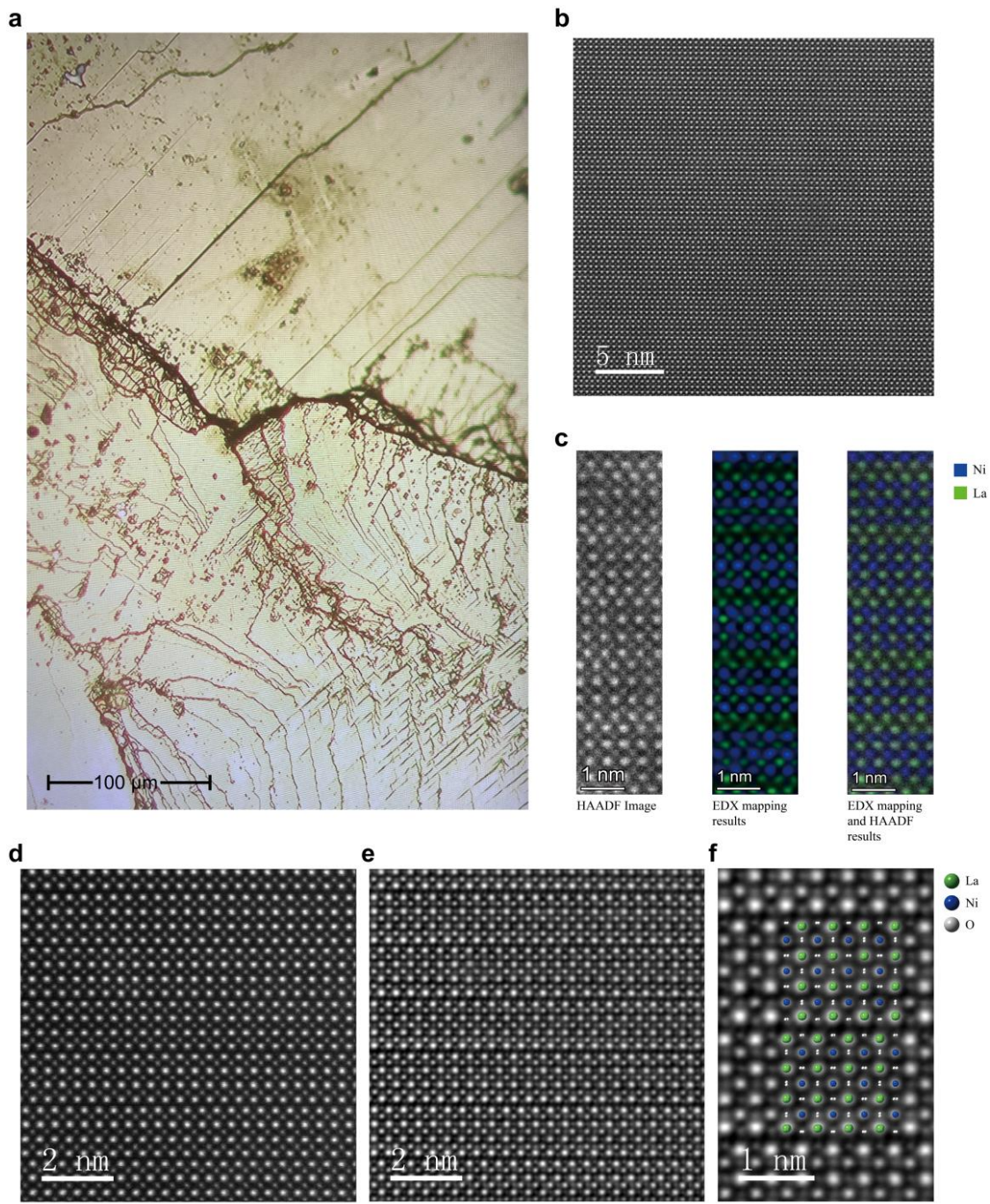

**Extended Data Fig. 6** | STEM-HAADF and integrated differential phase contrast (iDPC) measurements of $La_4Ni_3O_{10-\delta}$ crystals **a,** The cleavage plane of a $La_4Ni_3O_{10-\delta}$ single crystal. **b,** The STEM-HAADF image of a $La_4Ni_3O_{10-\delta}$ crystal along the [110] direction on the scale of 5 nm. **c,** Atomic EDX mapping results and STEM-HAADF image in the same region. **d,** The STEM-HAADF image of $La_4Ni_3O_{10-\delta}$ sample along the [110] direction on the scale of 2 nm. **e,** iDPC result in the same region and **f,** an enlarged view of the iDPC image. The oxygen atoms are clearly revealed through the iDPC imaging, aligning well with the crystal structure of $La_4Ni_3O_{10-\delta}$ determined by neutron diffraction and XRD measurements.

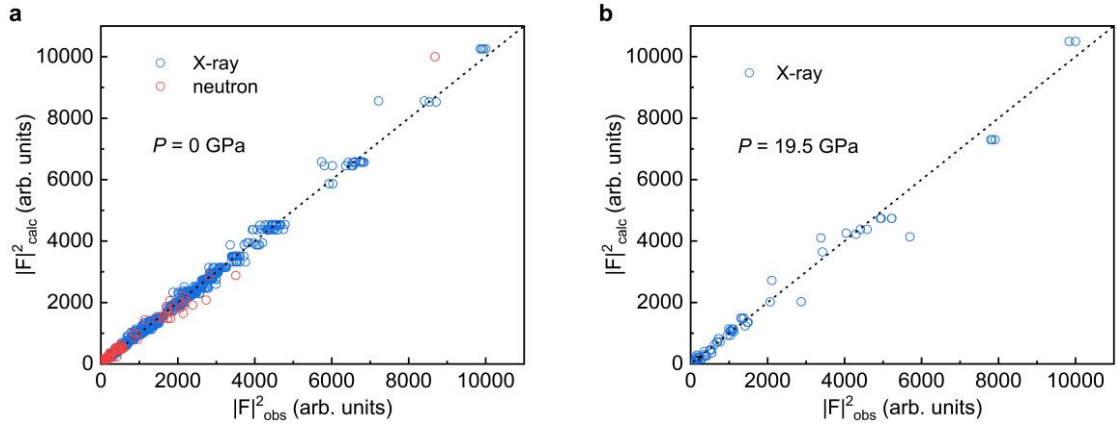

**Extended Data Fig. 7** | Single-crystal refinements using neutron and X-ray diffraction data for $La_4Ni_3O_{10-\delta}$. **a**, Combined single-crystal refinement results using neutron and X-ray diffraction data for $La_4Ni_3O_{10-\delta}$. The integrated intensities of Bragg reflections from a $La_4Ni_3O_{10-\delta}$ single crystal, collected at room temperature and ambient pressure, are plotted against their calculated counterparts. Details of the refined parameters are provided in Supplementary Table 1. **b**, Single-crystal refinements of X-ray diffraction data. Integrated intensities of the Bragg reflections collected on a $La_4Ni_3O_{10-\delta}$ single crystal at room temperature and 19.5 GPa are plotted against their calculated counterparts. The refinement parameters are summarized in Supplementary Table 2. The refinements suggest a structural phase transition from monoclinic $P2_1/a$ to tetragonal $I4/mmm$ in $La_4Ni_3O_{10-\delta}$ under pressure.

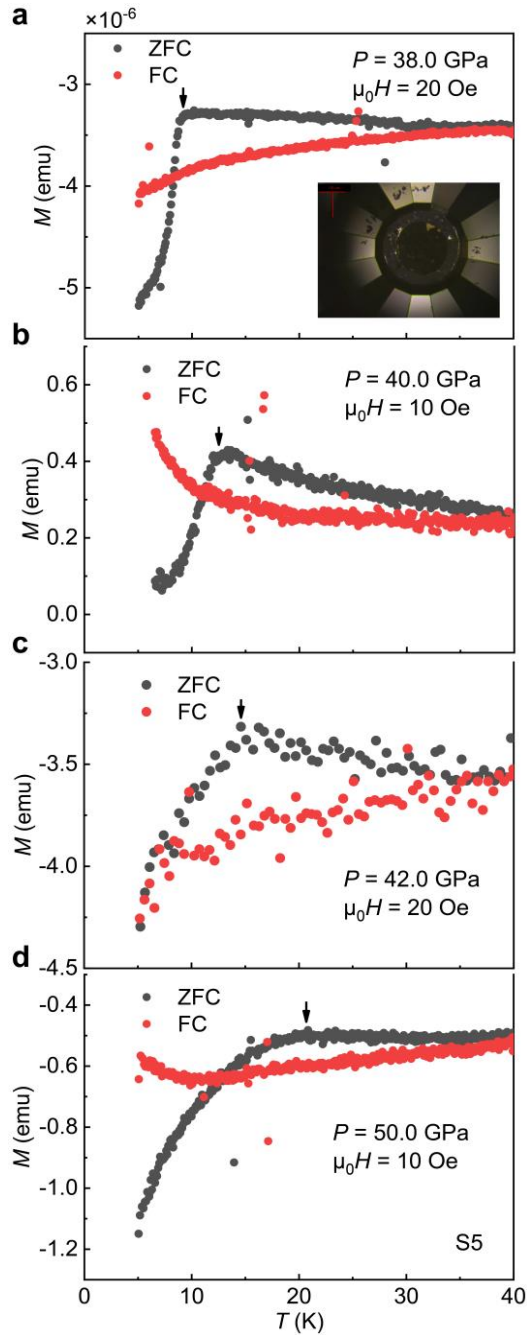

**Extended Data Fig. 8** | Temperature-dependent DC susceptibilities of sample 5 (S5) under various pressures. **a**, Temperature dependent magnetization curves at 38.0 GPa under a magnetic field of 20 Oe applied perpendicular to the *ab* plane using the zero-field-cooled (ZFC) and field-cooled (FC) modes. The inset shows the photo of the crystal in the nitrogen mini-DAC. **b**, 40.0 GPa and 10 Oe. **c**, 42.0 GPa and 20 Oe. **d**, 50.0 GPa and 10 Oe. The black arrows indicate the superconducting transition temperatures. The broadening of the superconducting transition with increasing pressure could be attributed to the gradual deterioration of hydrostaticity in the nitrogen pressure-transmitting medium. This broadening is less pronounced in a smaller sample (S6), which contains more pressure-transmitting medium and, consequently, exhibits improved hydrostaticity (Fig. 3f-i).

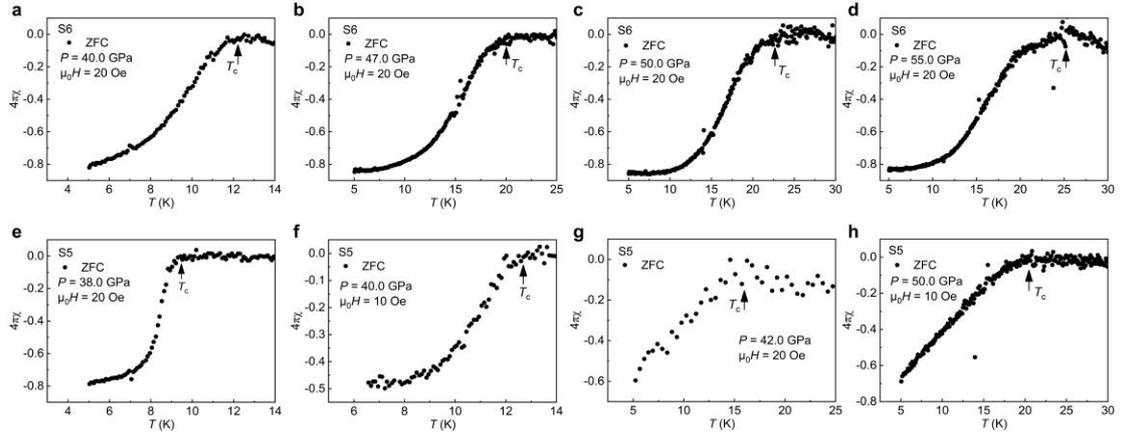

**Extended Data Fig. 9** | Temperature dependent superconducting volume fractions of $La_4Ni_3O_{10-\delta}$ single crystals determined through DC susceptibility measurements in Fig. 3 (S6) and Extended Data Fig. 8 (S5). Superconducting volume fractions for S6 under **a**, 40.0 GPa; **b**, 47.0 GPa; **c**, 50.0 GPa; **d**, 55.0 GPa. Superconducting volume fractions for S5 under **e**, 38.0 GPa; **f**, 40.0 GPa; **g**, 42.0 GPa; **h**, 50.0 GPa.

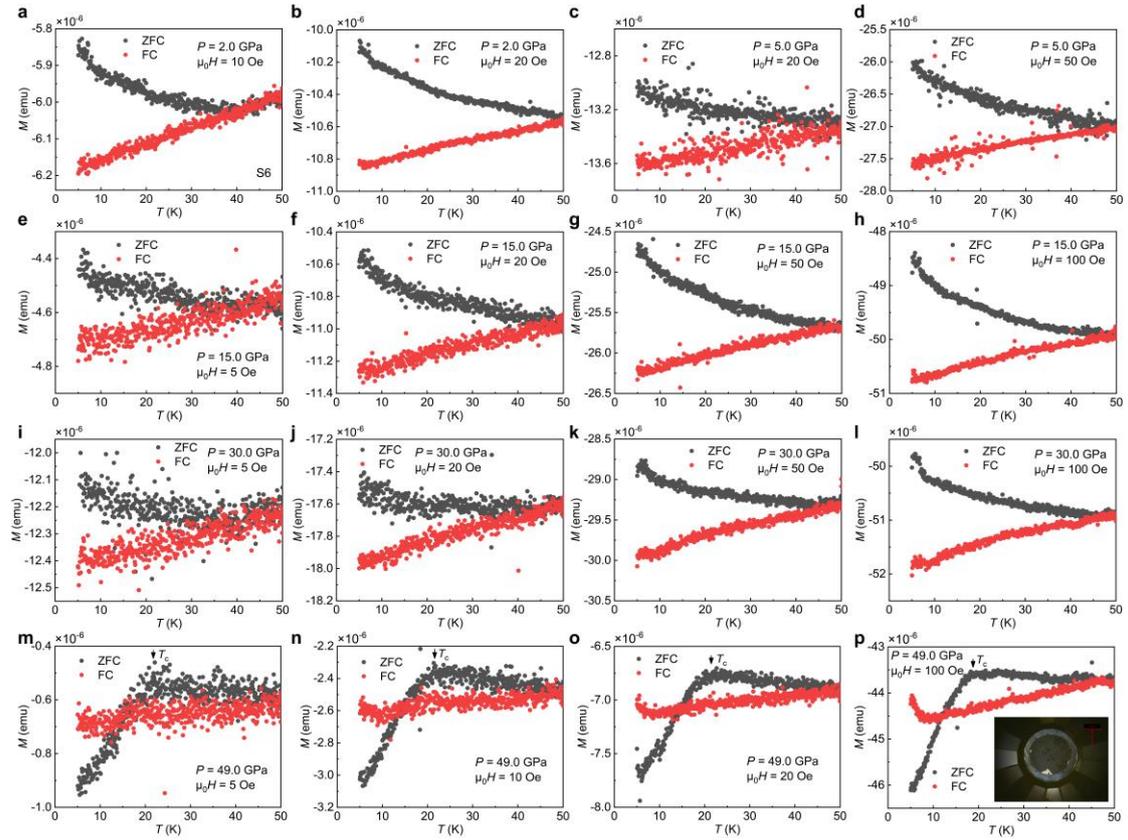

**Extended Data Fig. 10 |** Temperature-dependendent magnetization curves of sample 7 (S7) under various pressures under a magnetic field applied perpendicular to the *ab* plane using the zero-field-cooled (ZFC) and field-cooled (FC) modes. **a**, 2.0 GPa and 10 Oe **b**, 2.0 GPa 20 Oe. Temperature dependent magnetic moment at 5.0 GPa under magnetic fields of **c**, 5.0 GPa and 20 Oe and **d**, 5.0 GPa and 50 Oe. **e**, 15.0 GPa and 5 Oe; **f**, 15.0 GPa and 20 Oe; **g**, 15.0 GPa and 50 Oe; **h**, 15.0 GPa and 100 Oe. **i**, 30.0 GPa and 5 Oe; **j**, 30.0 GPa and 20 Oe; **k**, 30.0 GPa and 50 Oe; **l**, 30.0 GPa and 100 Oe. **m**, 49.0 GPa and 5 Oe; **n**, 49.0 GPa and 10 Oe; **o**, 49.0 GPa and 20 Oe; **p**, 49.0 GPa and 100 Oe. The inset shows the photo of the crystal in the mini-DAC.

# Supplementary Information for

# Superconductivity in pressurized trilayer La$_4$Ni$_3$O$_{10-\delta}$ single crystals

**Contents**

1. Supplementary Figures

**Fig. 1**: Chemical composition analysis using energy-dispersive X-ray (EDX) spectroscopy

**Fig. 2**: Specific heat of a La$_4$Ni$_3$O$_{10-\delta}$ single crystal

**Fig. 3**: Magnetization curves of La$_4$Ni$_3$O$_{10-\delta}$ for sample 7 (S7) below 300 K.

**Fig. 4**: Magnetization curves for sample 7 (S7) near $T_c$ under various magnetic fields

**Fig. 5**: Magnetization curves of empty mini-cells

**Fig. 6**: The surface morphology of La$_4$Ni$_3$O$_{10-\delta}$ single crystals

**Fig. 7**: The derivatives of the temperature-dependent resistance curves

**Fig. 8**: Comparison of *P2$_1$/a* and *Bmab* structural models

**Fig. 9**: Single-crystal X-ray and neutron diffraction measurements of La$_4$Ni$_3$O$_{10-\delta}$

2. Supplementary Tables

**Table 1**: Crystallographic parameters derived from a combined single-crystal refinement of neutron and X-ray diffractions at ambient condition

**Table 2**: Crystallographic parameters obtained from the single-crystal refinements of XRD data at 19.5 GPa

**Table 3**: Crystallographic parameters of La$_4$Ni$_3$O$_{10-\delta}$ obtained from Rietveld refinements of high-pressure synchrotron XRD measurements on powdered La$_4$Ni$_3$O$_{10-\delta}$ single crystals

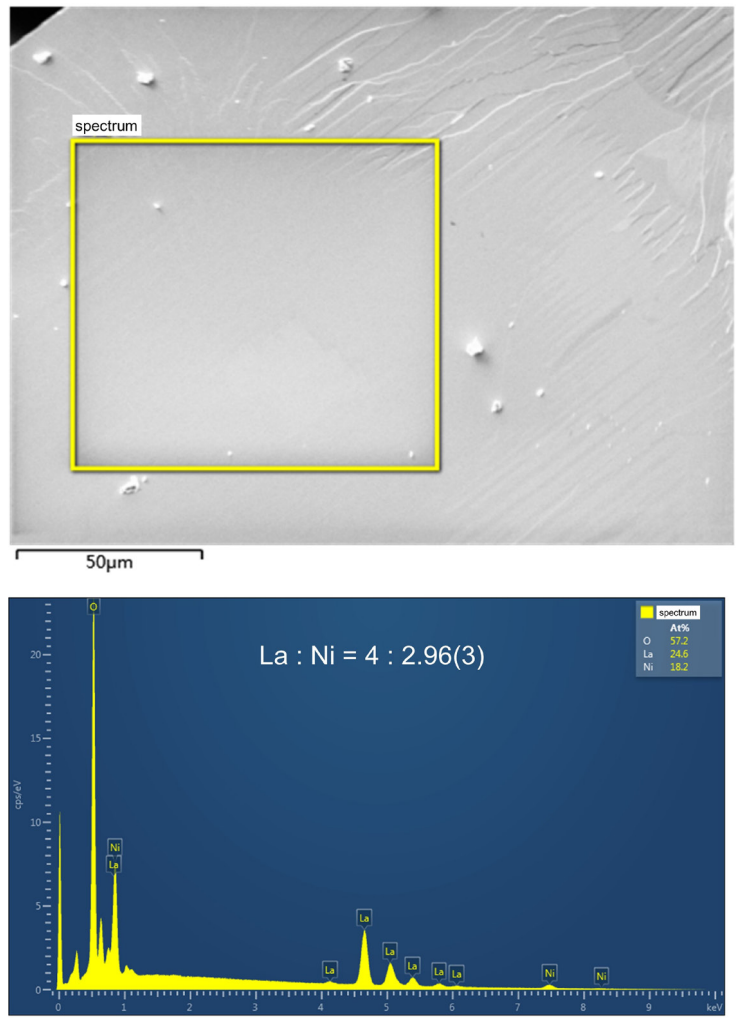

**Supplementary Fig. 1** | Chemical composition analysis using a scanning electron microscope (SEM) equipped with energy-dispersive X-ray (EDX) spectroscopy on a selected area of a $La_4Ni_3O_{10-\delta}$ single crystal. The analysis reveals a La to Ni atomic ratio of approximately 4 : 2.96(3), aligning with findings from neutron diffraction, X-ray diffraction refinements, and scanning transmission electron microscopy (STEM) investigations. This consistency underscores the accuracy of the material's stoichiometry as determined by multiple advanced imaging and analysis techniques.

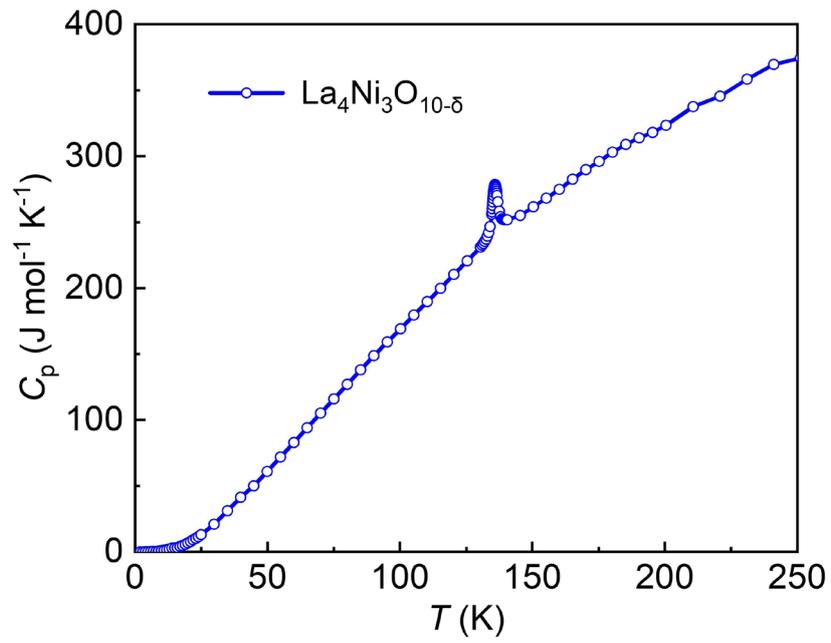

**Supplementary Fig. 2** | Temperature dependence of specific heat of a $La_4Ni_3O_{10-\delta}$ single crystal. A distinct feature corresponding to the spin density wave and charge density wave phase transition is clearly observable.

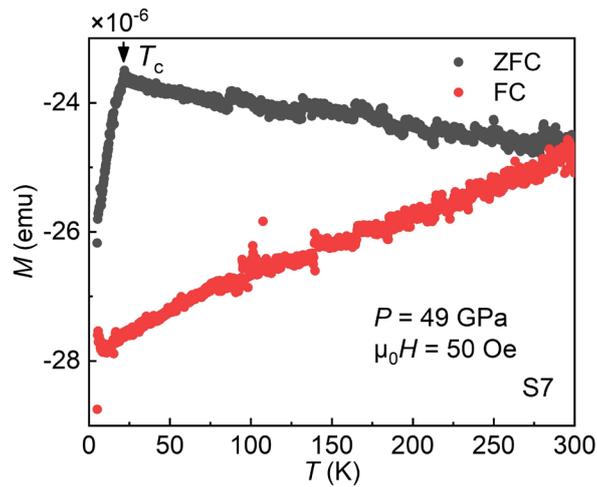

**Supplementary Fig. 3 │** Temperature dependent magnetization curves of $La_4Ni_3O_{10-\delta}$ under a magnetic field of 20 Oe applied perpendicular to the *ab* plane using the zero-field-cooled (ZFC) and field-cooled (FC) modes at 49.0 GPa for sample 7 (S7). A distinct superconducting diamagnetic response at $T_c$ is clearly observed in the ZFC curve. The normal state signals are predominantly influenced by the contribution from the pressure cell. The splitting of ZFC and FC curve in the normal state at the terminated temperature of 300 K is due to the presence of magnetic impurities from the pressure cell, which can be seen clearly in the empty cell runs at different temperatures (Supplementary Figure 5, ref. 49).

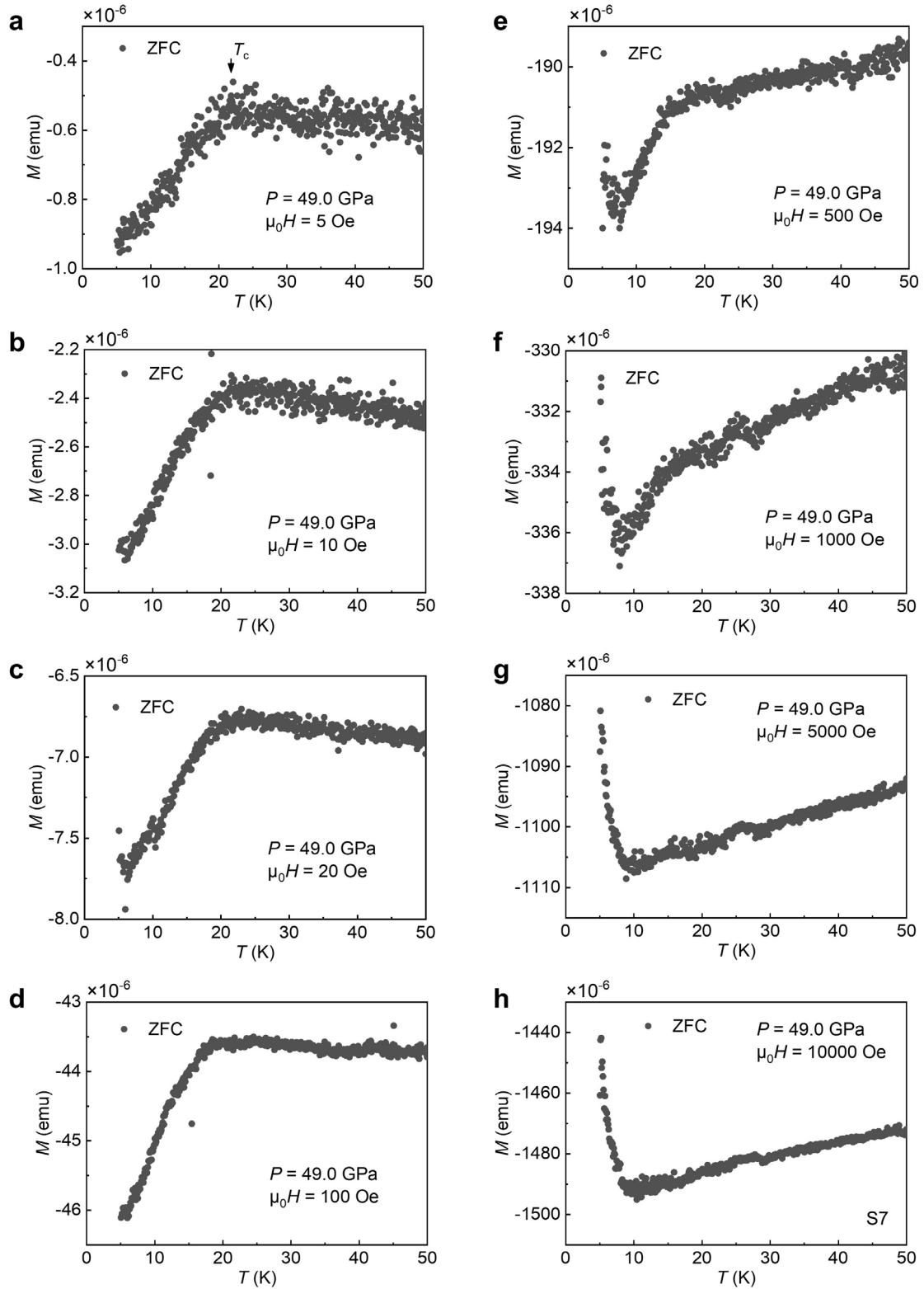

**Supplementary Fig. 4** | Temperature dependent magnetization curves for sample 7 (S7) at 49.0 GPa under the external magnetic fields of **a**, 5 Oe; **b**, 10 Oe; **c**, 20 Oe; **d**, 100 Oe; **e**, 500 Oe; **f**, 1000 Oe; **g**, 5000 Oe; **h**, 10000 Oe. Superconducting diamagnetic signals are distinctly visible at lower fields. However, as the magnetic field increases, the background magnetic signal from the pressure cell increases rapidly, making it difficult to discern the superconducting diamagnetic signal at fields above 10000 Oe.

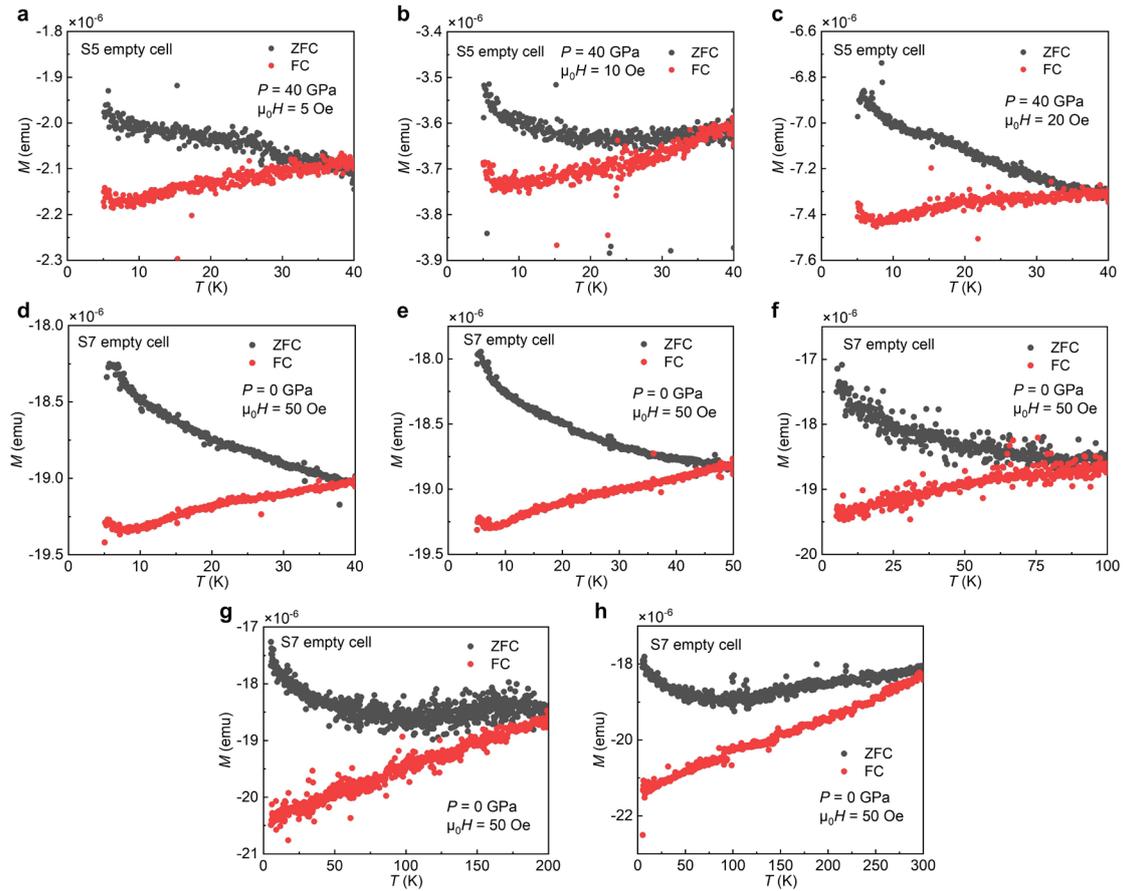

**Supplementary Fig. 5** | Temperature dependent magnetization curves of the empty mini-cell used for sample 5 (S5) run at 40 GPa and **a**, 5 Oe; **b**, 10 Oe; **c** 20 Oe. Temperature dependent magnetization curves of the empty cell used for sample 7 (S7) run at ambient pressure and 50 Oe within **d**, 5-40 K; **e**, 5-50 K; **f**, 5-100 K; **g**, 5-200 K; **h**, 5-300 K. The magnetic fields are applied perpendicular to the *ab* plane using the zero-field-cooled (ZFC) and field-cooled (FC) modes. The pressure cell for two different runs may have slightly different amount of rhenium used as a gasket.

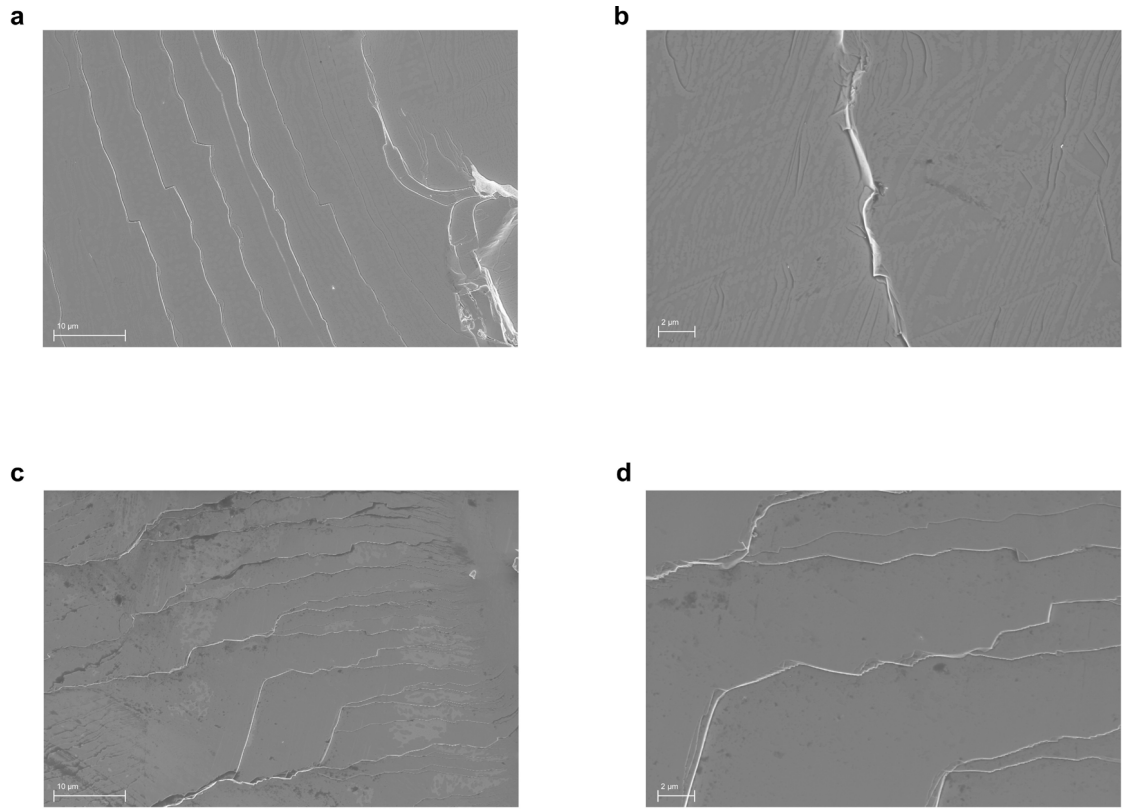

**Supplementary Fig. 6** | The surface morphology of the cleavage plane of $La_4Ni_3O_{10-\delta}$ single crystals observed through scanning electron microscopy (SEM). SEM images **a**, and **c**, were obtained with a scale bar of 10 μm. Zoom-in SEM images **b**, and **d**, were obtained with a scale bar of 2 μm. Across all images, no detectable impurities were found, affirming the purity of the single crystals utilized in high-pressure resistance and susceptibility measurements.

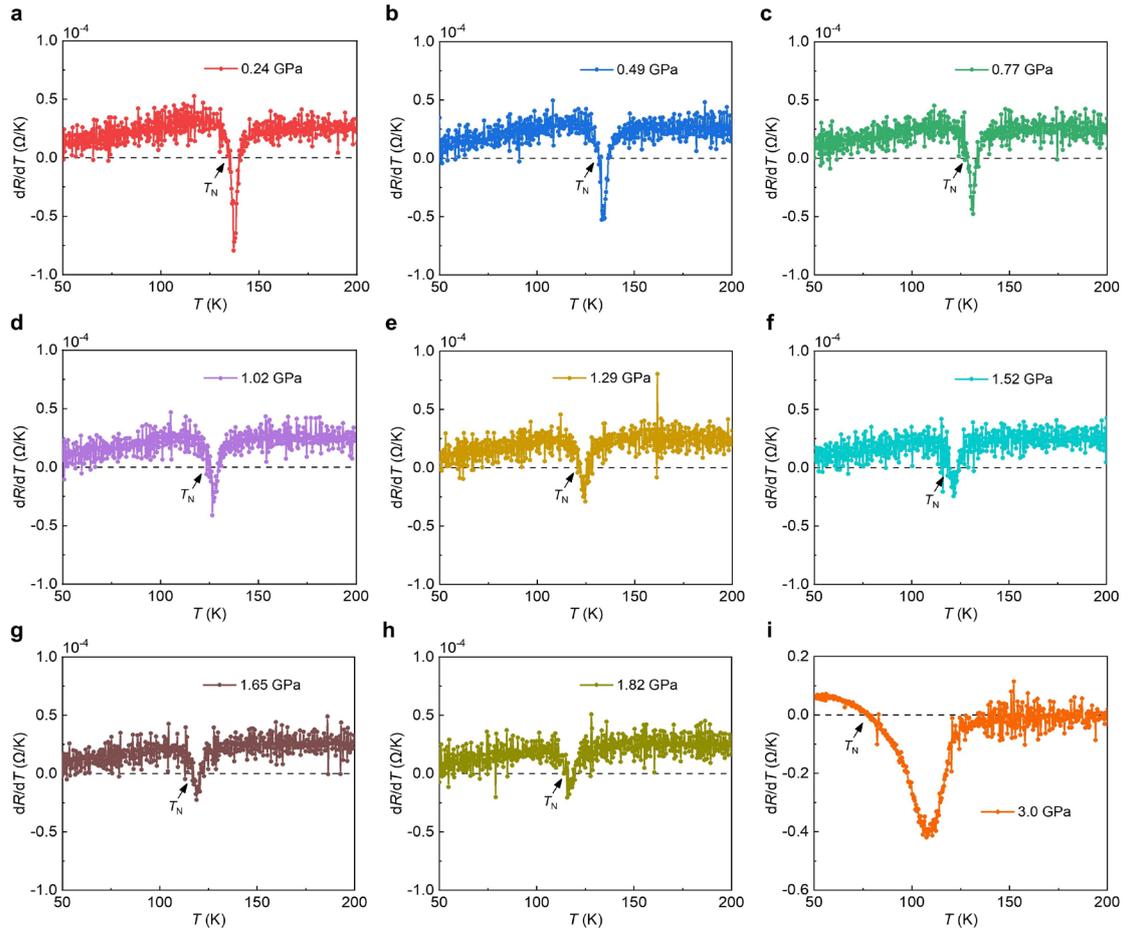

**Supplementary Fig. 7**│ The derivatives of the temperature-dependent resistance curves (dR/dT) derived from Figs. 3a and 3b under varying pressures. **a**, 0.24 GPa; **b**, 0.49 GPa; **c**, 0.77 GPa; **d**, 1.02 GPa; **e**, 1.29 GPa; **f**, 1.52 GPa. **g**, 1.65 GPa. **h**, 1.82 GPa. **i**, 3.0 GPa. The Néel temperature ($T_N$) from resistivity measurements is identified at the maximum of the resistance anomaly, where the temperature derivative dR/dT equals zero. Resistance anomalies at 11 GPa and 15.5 GPa in Fig. 3b, which are influenced by the onset of superconductivity, complicate accurate estimation of $T_N$ and therefore are not included.

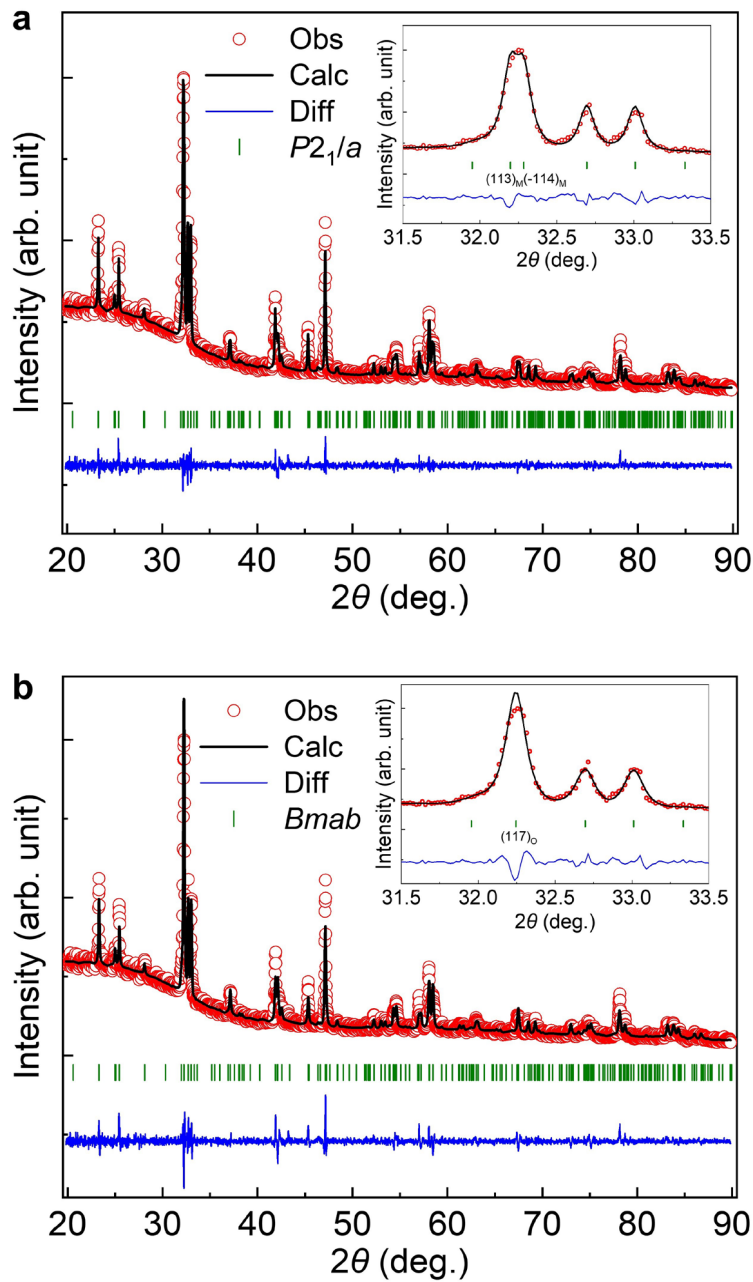

**Supplementary Fig. 8 | Comparison of *P2₁/a* and *Bmab* structural models at ambient condition. a**, Rietveld refinement of a lab-based XRD pattern for powdered La$_4$Ni$_3$O$_{10-\delta}$ single crystals at ambient pressure and room temperature. Black solid lines represent the fit to the experimental data, shown by red circles. Blue solid lines indicate the residual difference between observed and calculated intensities. Short vertical green bars mark the positions of expected Bragg peaks. This dataset fits well with the *P2₁/a* space group (GOF = 2.0, R$_p$ = 4.12, R$_{wp}$ = 5.46). The inset shows the details near 32.2°, where the (1 1 3)$_M$ and (-1 1 4)$_M$ peaks are captured by the *P2₁/a* model. **b**, Rietveld refinement for the *Bmab* structure using the same XRD data (GOF = 2.4, R$_p$ = 4.74, R$_{wp}$ = 6.62). The inset shows that the *Bmab* model fails to resolve the double peak structure near 32.2°, instead presenting a single peak of (1 1 7)$_O$ which suggests a less accurate representation of the underlying crystal structure. The measurements were carried out on a Bruker D8 Discover diffractometer, utilizing Cu Kα radiation (λ=1.5406 Å).

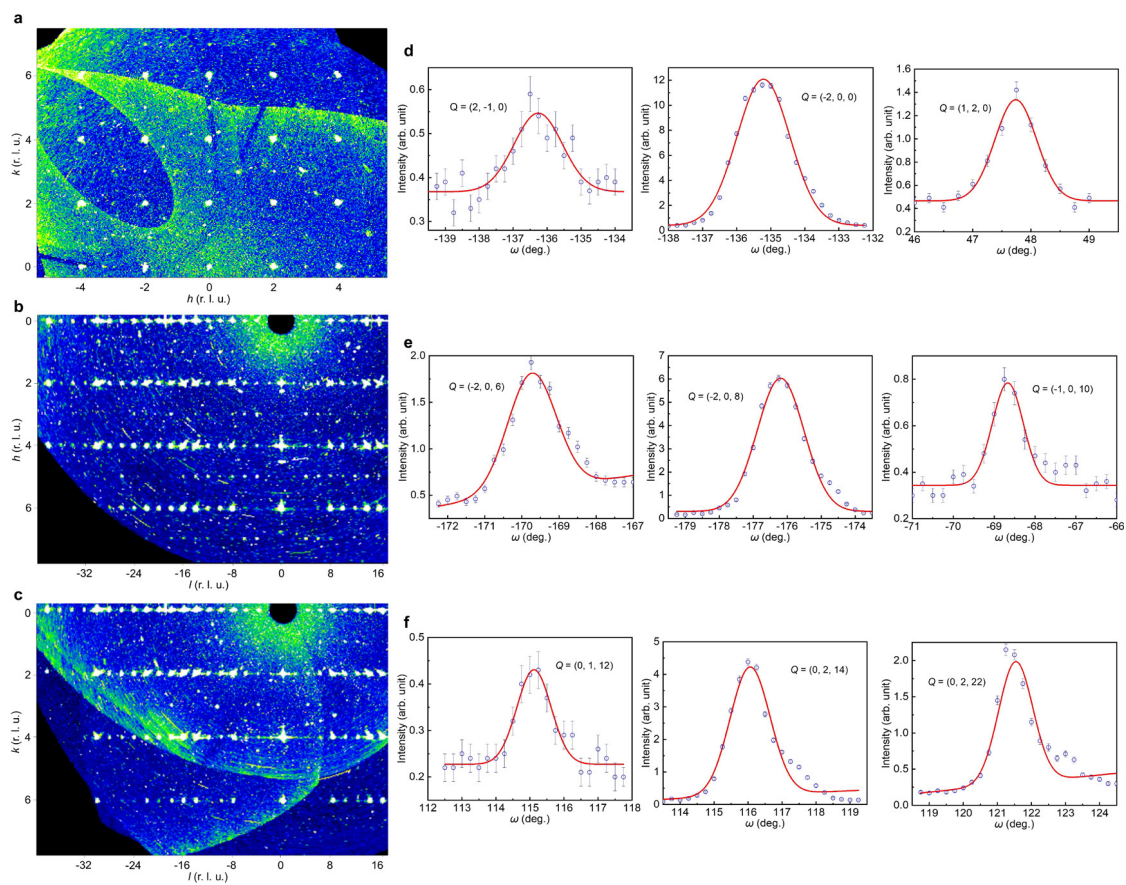

**Supplementary Fig. 9** | Single-crystal X-ray and neutron diffraction measurements of $La_4Ni_3O_{10-\delta}$ at ambient pressure and room temperature. For clarity in visualizing the main Bragg peaks corresponding to both the $P2_1/a$ and $Bmab$ structures, we have plotted the data based on the $P2_1/a$ structure with $Z=4$, having unit cell dimensions $a = 5.4164$ Å, $b = 5.4623$ Å, and $c = 27.985$ Å. **a**, ($h\ k\ 8$) reciprocal plane reconstructed from X-ray diffraction measurements. In addition to the strong Bragg peaks corresponding to the $Bmab$ structure, which occur when both $h$ and $k$ are even, additional weaker peaks (when $h\ k$ are integers) due to the monoclinic distortion inherent to the $P2_1/a$ structure are also observed. This indicates $Bmab$ space group cannot accurately represent the structure; **b**, ($h\ 0\ l$) reconstructed reciprocal plane; **c**, ($0\ k\ l$) reconstructed reciprocal plane; **d**, Representative neutron diffraction rocking curves of the ($h\ k\ 0$) Bragg peaks; **e**, ($h\ 0\ l$) Bragg peaks; **f**, ($0\ k\ l$) Bragg peaks. The comparisons between the measured and calculated intensities of these Bragg peaks can be found in Extended Data Fig. 7a. The X-ray diffraction measurements were conducted using a lab-based Bruker D8 Venture diffractometer, while the neutron diffraction measurements utilized the HB-3A four-circle single-crystal neutron diffractometer. Error bars, 1 s.d.

**Supplementary Table 1** Crystallographic parameters derived from a combined single-crystal refinement of neutron and X-ray diffractions on $La_4Ni_3O_{9.96}$ single crystals, at ambient pressure and room temperature. All O occupancies are constrained between 0 and 1.

| | $P2_1/a$ (Z = 2) | | |
|---|---|---|---|
| | a = 5.4164(3) Å | b = 5.4623(3) Å | c = 14.2279(2) Å |
| | α = 90° | β = 100.75° | γ = 90° |
| | **Atomic positions** | | |
| **Name (site)** | *x* | *y* | *z* | *Occ* |
| La1 (4e) | 0.803(1) | 0.508(1) | 0.103(1) | 1.00 |
| La2 (4e) | 0.939(1) | 0.500(1) | 0.364(1) | 1.00 |
| Ni1 (2d) | 0.5 | 0.5 | 0.5 | 1.00 |
| Ni2 (4e) | 0.862(1) | 0.002(1) | 0.222(1) | 1.00 |
| O1 (4e) | 0.786(5) | 0.967 (2) | 0.068(1) | 1.00 |
| O2 (4e) | 0.066(4) | 0.954(2) | 0.636(1) | 1.00 |
| O3 (4e) | 0.112(4) | 0.742(5) | 0.231(1) | 1.00 |
| O4 (4e) | 0.113(4) | 0.248(6) | 0.211(1) | 0.98(2) |
| O5 (4e) | 0.771(4) | 0.723(4) | 0.486(1) | 1.00 |

$R_F$ = 2.99, $R_{F2}$ = 4.10, $R_{wF2}$ = 7.32

**Supplementary Table 2** Crystallographic parameters obtained from the single-crystal refinement of XRD data in La$_4$Ni$_3$O$_{10-\delta}$, conducted at 19.5 GPa and room temperature. A methanol-ethanol-water mixture in the ratio of 16:3:1 was used as the pressure-transmitting medium for the measurement.

| *I4/mmm* (Z = 2) | | | |
|---|---|---|---|
| $a$ = 3.73(1) Å | $b$ = 3.73(1) Å | $c$ = 27.2(1) Å | |
| $\alpha$ = 90° | $\beta$ = 90° | $\gamma$ = 90° | |
| **Atomic positions** | | | |
| **Name (site)** | $x$ | $y$ | $z$ |
| La1 (4e) | 1.0 | 1.0 | 0.432(1) |
| La2 (4e) | 1.0 | 1.0 | 0.302(1) |
| Ni1 (2a) | 0.5 | 0.5 | 0.5 |
| Ni2 (4e) | 0.5 | 0.5 | 0.358(1) |
| O1 (8g) | 1 | 0.5 | 0.363(1) |
| O2 (4e) | 0.5 | 0.5 | 0.392(8) |
| O3 (4c) | 0 | 0.5 | 0.5 |
| O4 (4e) | 0.5 | 0.5 | 0.281(2) |

$R_F$ = 6.25, $R_{F2}$ = 9.28, $R_{wF2}$ = 14.8

**Supplementary Table 3** Crystallographic parameters of pressurized $La_4Ni_3O_{10-\delta}$ obtained from Rietveld refinements of high-pressure synchrotron XRD. Helium was used as the pressure-transmitting medium for the measurement.

| Pressure (GPa) | a(Å) | b(Å) | c(Å) | β(°) | GOF | $R_p$ (%) | $R_{wp}$ (%) | Space group |
|---|---|---|---|---|---|---|---|---|
| 0 | 5.4142(4) | 5.4647(4) | 14.220(1) | 100.66 | 0.99 | 5.66 | 6.88 | $P2_1/a$ |
| 2.5 | 5.3852(3) | 5.4321(3) | 14.147(1) | 100.71 | 0.72 | 8.03 | 7.11 | $P2_1/a$ |
| 3.5 | 5.3769(40) | 5.4111(39) | 14.127(12) | 100.67 | 0.22 | 9.69 | 9.30 | $P2_1/a$ |
| 5.0 | 5.3675(8) | 5.3962(8) | 14.100(3) | 100.75 | 0.79 | 9.19 | 7.86 | $P2_1/a$ |
| 6.5 | 5.3478(4) | 5.3824(4) | 14.085(1) | 100.80 | 1.2 | 10.9 | 10.4 | $P2_1/a$ |
| 8.3 | 5.3320(6) | 5.3522(8) | 13.983(2) | 100.67 | 1.0 | 8.68 | 8.63 | $P2_1/a$ |
| 10.0 | 5.3250(5) | 5.3406(5) | 13.972(1) | 100.76 | 0.93 | 10.4 | 8.91 | $P2_1/a$ |
| 11.7 | 5.2950(4) | 5.3265(5) | 13.910(1) | 100.66 | 0.72 | 8.00 | 6.91 | $P2_1/a$ |
| 13.7 | 3.7485(6) | 3.7485(6) | 27.346(5) | 90 | 1.0 | 7.15 | 7.55 | $I4/mmm$ |
| 15.2 | 3.7310(1) | 3.7310(1) | 27.207(1) | 90 | 0.88 | 7.13 | 6.91 | $I4/mmm$ |
| 17.2 | 3.7245(1) | 3.7245(1) | 27.130(1) | 90 | 0.85 | 7.46 | 7.19 | $I4/mmm$ |
| 18.7 | 3.7177(6) | 3.7177(6) | 27.053(5) | 90 | 0.72 | 7.13 | 6.53 | $I4/mmm$ |
| 21.6 | 3.7121(1) | 3.7121(1) | 27.023(2) | 90 | 0.61 | 6.50 | 5.86 | $I4/mmm$ |
| 24.0 | 3.6934(1) | 3.6934(1) | 26.880(1) | 90 | 0.78 | 5.92 | 5.94 | $I4/mmm$ |
| 27.1 | 3.6879(1) | 3.6879(1) | 26.854(2) | 90 | 0.75 | 7.15 | 6.92 | $I4/mmm$ |
| 30.0 | 3.6716(1) | 3.6716(1) | 26.704(2) | 90 | 0.86 | 7.82 | 7.90 | $I4/mmm$ |
| 34.0 | 3.6602(1) | 3.6602(1) | 26.609(2) | 90 | 0.68 | 7.00 | 6.44 | $I4/mmm$ |
| 38.0 | 3.6487(2) | 3.6487(1) | 26.534(2) | 90 | 1.3 | 9.1 | 8.9 | $I4/mmm$ |
| 42.0 | 3.6319(1) | 3.6319(1) | 26.452(2) | 90 | 1.1 | 8.99 | 8.75 | $I4/mmm$ |
| 48.0 | 3.6169(1) | 3.6169(1) | 26.316(1) | 90 | 0.94 | 8.14 | 7.97 | $I4/mmm$ |
| 52.0 | 3.6068(1) | 3.6068(1) | 26.243(1) | 90 | 0.88 | 7.82 | 7.51 | $I4/mmm$ |
| 56.0 | 3.5957(1) | 3.5957(1) | 26.186(2) | 90 | 0.95 | 6.62 | 6.83 | $I4/mmm$ |